\documentclass[pra,twocolumn,floatfix]{revtex4-1}

\usepackage[utf8]{inputenc}
\usepackage{xcolor}
\usepackage[backref=none]{hyperref}
\hypersetup{
    colorlinks=true,          
    linkcolor=blue,           
    citecolor=blue,           
    urlcolor=olive,           
}

\usepackage{amsmath}
\usepackage{amssymb}
\usepackage{multirow}
\usepackage{graphicx}
\graphicspath{ {images/} }
\usepackage{array}
\usepackage{upgreek}
\usepackage{mathtools}
\usepackage{cuted}

\usepackage[switch]{lineno}

\newcommand\yy[1]{\color{black}#1}
\newcommand\lk[1]{\color{black}#1}

\makeatletter

\begin{document}

\title{Efimov scenario for overlapping narrow Feshbach resonances}

\author{Yaakov Yudkin}
\author{Lev Khaykovich}

\affiliation{Department of Physics, QUEST Center and Institute of Nanotechnology and Advanced Materials, Bar-Ilan University, Ramat-Gan 5290002, Israel}

\date{\today}

\begin{abstract}

While Efimov physics in ultracold atoms is usually modeled with an isolated Feshbach resonance many real world resonances appear in close vicinity to each other and are therefore overlapping.
Here we derive a realistic model based on the mutual coupling of an open channel and two closed molecular channels while neglecting short-range physics as permitted by the narrow character of the considered resonances.
The model is applied to three distinct scenarios with experimental relevance.
{\lk We show that the effect of overlapping resonances is manifested most strikingly at a narrow resonance in whose vicinity there is a slightly narrower one.
In this system the Efimov ground state extends not only over the scattering length zero crossing between the two resonances but also over the pole of the second resonance to finally meet the dissociation threshold below it.
In the opposite scenario, when a narrow resonance is considered in the vicinity of a slightly broader one, we observe that the Efimov features are pushed to lower binding energies and smaller scattering lengths by a significant factor facilitating their experimental investigation.
Both scenarios are referenced to the case of two narrow resonances which are far enough away from each other to be effectively decoupled.
In this case the two-channel model results are recovered.}
Finally, we analyze the rich excitation spectrum of the system and construct and explain its {\yy {\yy nodal pattern}}.

\end{abstract}

\maketitle

\section{Introduction}


Tunablility of the $s$-wave scattering length $a$ via a magnetic Feshbach resonance is at the heart of recent studies of few-body physics in ultracold atoms~\cite{Greene17,Naidon17,D'Incao18}.
Conceptually, a Feshbach resonance can be understood within a simple two-channel model: it occurs when incoming atoms in an open channel are coupled to an almost degenerate bound state in a closed channel~\cite{Chin10}.
Loosely speaking, one differentiates between two types of Feshbach resonances quantified by the dimensionless resonance strength parameter $s_{res}$.
A broad resonance ($s_{res}\gg1$) arises from a strong coupling to the bound state.
The scattering amplitude is largely dominated by $a$, which is on the order of the van der Waals length $r_{vdW}$ (the range of the interaction potential) away from collisional resonances.
On the contrary, a narrow resonance ($s_{res}\ll1$) arises from a weak coupling.
In this case the effective range $r_e$, in addition to $a$, determines the scattering amplitude at a given magnetic field.
As a new length scale associated with the narrow resonance one defines $R^\star=-r_e(B_{\text{res}})/2>0$, where $r_e(B_{\text{res}})$ is the value of $r_e$ at the resonance position ($|a|\rightarrow\infty$).
$R^\star$ is related to $s_{res}$ via $R^\star=\bar{a}/s_{res}$ where $\bar{a}=[4\pi/\Gamma(1/4)^2]r_{vdW}$ is the mean scattering length~\cite{Chin10}.
Thus, for a broad (narrow) resonance $R^\star\ll r_{vdw}$ ($R^\star\gg r_{vdw}$) is satisfied~\cite{broadVSnarrow}.

It is clear, however, that the description of a real world scattering system, which in general is a multi-channel problem, within the framework of a two-channel model is an approximation and should be applied with caution.
It is worth noting that nearly all atomic species used in experiments exhibit multiple, often overlapping, Feshbach resonances.
Extreme examples include recently studied cold molecules with their complex internal structure~\cite{Yang19} and heavy lanthanide species where a dense and chaotic spectrum of Feshbach resonances has been reported~\cite{Frisch14}.

Ironically, simple analytical or semi-analytical expressions have been developed to describe the scattering length with great precision even in the case of a diverging number of scattering channels~\cite{Lange09,Jachymski13,Mehta18}.
This has lured few-body physics treatments to consider an isolated Feshbach resonance a good approximation for calculating various properties of few-body systems such as the energy levels of Efimov trimers~\cite{Efimov70}.
The latter, which is the main focus of this paper, form in a system with three atoms when the pairwise interactions exceed the relevant length scale, so $a\gg r_{vdW}$ ($a\gg R^\star$) for a broad (narrow) resonance.
Efimov physics has been studied extensively in the recent decade, theorerically and experimentally~\cite{Greene17,Naidon17,D'Incao18}, in the vicinity of both isolated broad~\cite{Gross09,Berninger11,Tung14,Pires14} and, more recently, narrow Feshbach resonances~\cite{Roy13,Johansen17,Chapurin19,Xie20}.
Many of these studies were performed in the vicinity of overlapping resonances~\cite{Gross09,Berninger11,Tung14,Pires14,Roy13,Johansen17}, but the theoretical treatment rarely goes beyond an isolated resonance.
For a few exception see Refs.~\cite{Wang14,Kato17}.

Recent efforts to fully incorporate the multi-channel character of two-body interactions suggest properly including the hyperfine structure of the real atomic system~\cite{Chapurin19,Xie20,Secker20}.
Although this is arguably the most comprehensive approach, it comes at the expense of heavy numerical calculations and the absence of direct relations between the microscopic parameters of the theory and the macroscopic experimental observables.
Here, in contrast, we consider the simplest way to deal with a consequence of the multi-channel character of two-body interactions, namely the existence of two {\it overlapping} Feshbach resonances.
We generalize the two-channel model, which is suitable for isolated narrow Feshbach resonances~\cite{Petrov04}, to two overlapping resonances by including a second closed channel with an independently tunable bound state.
For completeness an inter-molecular coupling between the two closed channels is incorporated.
We develop a protocol to fix all model parameters in the two-body sector and compute the Efimov spectrum without any adjustable parameters.
{\lk In three-body sector we identify unique features related to the addition of a second closed channel and discuss their experimental implications.}

The paper is organized as follows.
We begin in Sec.~\ref{sec:3-channel phenomenology} by examining the expected phenomenology of the three-channel model.
In Sec.~\ref{sec:3-channel derivation} we state the three-channel model Hamiltonian and derive equations for the scattering amplitude, dimer binding energy and trimer binding energy.
The model is applied to three distinct systems in Sec.~\ref{sec:model systems} and their experimental relevance is addressed.
The model's bare parameters are further discussed in Sec.~\ref{sec:Lambda_bb} and the trimer eigen functions are analyzed in Sec.~\ref{sec:eigen functions}.
We conclude in Sec.~\ref{sec:conclusions} with possible extensions of the three-channel model.

\section{Phenomenology of the Three-Channel Model}
\label{sec:3-channel phenomenology}

\begin{figure}[b]
\centering
\includegraphics[width=1.\linewidth]{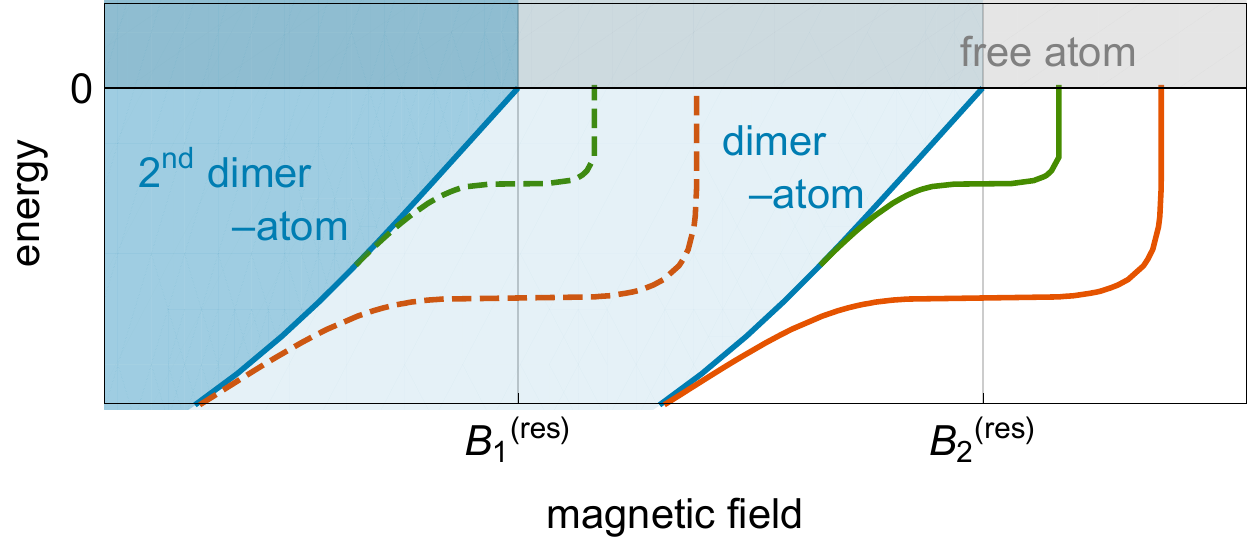}
\caption{\label{fig:illustration}
Schematic representation of the three-body sector of overlapping resonances.
The two resonance positions are labeled $B_{1,2}^{(\text{res})}$.
The binding energies of the dimers (blue curves) as well as the ground (orange) and excited (green) state trimers are plotted.
The regions of the free atom continuum (gray, positive energy) and the two dimer-atom continua (shades of blue) are indicated.
The trimers associated with $B_{1}^{(\text{res})}$ are embedded in the dimer-atom continuum due to the resonance at $B_{2}^{(\text{res})}$.
}
\end{figure}

The two-channel model is very successful at reproducing the basic phenomenology of a narrow and isolated Feshbach resonance.
It captures both the two- and the three-body sector~\cite{Petrov04}.
In its most fundamental form one considers a featureless open channel {\yy (i.e. no background scattering)} and, coupled to it, a closed (molecular) channel detuned from the open channel by a magnetic field dependent binding energy.
For completeness, this model is reviewed in Appendix~\ref{app:2-channel}.
It has been applied to various aspects of bosonic~\cite{Gogolin08,Pricoupenko10,Jona-Lasino10,Pricoupenko11,Levinsen15} and fermionic~\cite{Castin06,Jona-Lasinio08,Nishida12,Nishida15,Yi15,Pierce19} scenarios and may be generalized to include background scattering in the open channel~\cite{Werner09} and to hetero-nuclear systems~\cite{Castin10,Mora11}.

Here we take a different route and generalize the model to three channels by adding a second molecular channel.
The coupling of the two molecular channels to the free atom continuum gives rise to two scattering resonances and two two-body bound states which we call dimers.
Note the distinction between a molecular state, which is a bare state of the (non-interacting) Hamiltonian, and a dimer, which is the two-body bound eigen state of the full Hamiltonian.
In the three-body sector, which is schematically illustrated in Fig.~\ref{fig:illustration}, there are three types of continua.
In addition to the free atom continuum there are two different dimer-atom continua -- one above each dimer.
Around each of the resonances and below the respective dimers we expect three-body bound states, i.e. Efimov trimers.
Note though that only the trimers associated with the higher resonance (higher magnetic field, $B_2^{(\text{res})}$ in Fig.~\ref{fig:illustration}) are true bound states.
The $B_{1}^{(\text{res})}$ trimers coexist with a continuum of dimer-atom states and are therefore not stable bound states.
The question of their existence depends on their lifetime and they might be manifested as dimer-atom collision resonances~\cite{Jona-Lasinio08}. 
This subject is, however, beyond the scope of the present discussion in which only true bound states are considered. 

\section{Derivation of the Three-Channel Model}
\label{sec:3-channel derivation}

The following three-channel Hamiltonian is the most general extension of the two-channel model.
We use it to formulate equations for the scattering amplitude and the dimer binding energy (two-body sector).
We examine the two-body sector and relate the bare parameters to physical observables before moving on to the three-body sector where the equation for Efimov bound states is presented and discussed.

\subsection{Three-Channel Hamiltonian}

The full Hamiltonian of the three-channel model is $\hat{H}=\hat{H}_0+\hat{H}_{int}$, where $\hat{H}_{int}=\hat{H}_{1}+\hat{H}_{2}+\hat{H}_{12}$.
The first term is the bare Hamiltonian of all three channels
\begin{multline}
\hat{H}_{0}=\int\frac{d^{3}k}{\left(2\pi\right)^{3}}\bigg[\frac{\hbar^{2}k^{2}}{2m}\hat{a}_{\vec{k}}^{\dagger}\hat{a}_{\vec{k}}+\left(E_{b,1}+\frac{\hbar^{2}k^{2}}{4m}\right)\hat{b}_{\vec{k}}^{\dagger}\hat{b}_{\vec{k}} \\
+\left(E_{b,2}+\frac{\hbar^{2}k^{2}}{4m}\right)\hat{c}_{\vec{k}}^{\dagger}\hat{c}_{\vec{k}}\bigg],
\end{multline}
where $\hat{a}_{\vec{k}}$ annihilates free particles and $\hat{b}_{\vec{k}}$ ($\hat{c}_{\vec{k}}$) annihilates molecules in the first (second) molecular channel with bare molecular energy $E_{b,1}$ ($E_{b,2}$).
We assume the latter to be an affine function of the magnetic field: $E_{b,i}=\mu_i\left(B_i-B\right)$, where $\mu_i$ is the magnetic moment of the $i$-th molecular channel with respect to the free atom continuum and $B_i$ is the bare resonance position.
Without loss of generality we assume $B_1<B_2$.
Both molecular channels are coupled to the open channel via
\begin{multline}
\hat{H}_{1}=\Lambda_1\int\frac{d^{3}k}{\left(2\pi\right)^{3}}\int\frac{d^{3}q}{\left(2\pi\right)^{3}} \\
\left[\hat{b}_{\vec{k}}^{\dagger}\hat{a}_{\vec{q}+\frac{\vec{k}}{2}}\hat{a}_{-\vec{q}+\frac{\vec{k}}{2}}+\hat{a}_{-\vec{q}+\frac{\vec{k}}{2}}^{\dagger}\hat{a}_{\vec{q}+\frac{\vec{k}}{2}}^{\dagger}\hat{b}_{\vec{k}}\right]
\end{multline}
and
\begin{multline}
\hat{H}_{2}=\Lambda_2\int\frac{d^{3}k}{\left(2\pi\right)^{3}}\int\frac{d^{3}q}{\left(2\pi\right)^{3}} \\
\left[\hat{c}_{\vec{k}}^{\dagger}\hat{a}_{\vec{q}+\frac{\vec{k}}{2}}\hat{a}_{-\vec{q}+\frac{\vec{k}}{2}}+\hat{a}_{-\vec{q}+\frac{\vec{k}}{2}}^{\dagger}\hat{a}_{\vec{q}+\frac{\vec{k}}{2}}^{\dagger}\hat{c}_{\vec{k}}\right].
\end{multline}
In addition, the two molecular channels are coupled to each other via
\begin{equation}
\hat{H}_{12}=\Lambda_{12}\int\frac{d^{3}k}{\left(2\pi\right)^{3}}\left[\hat{c}_{\vec{k}}^{\dagger}\hat{b}_{\vec{k}}+\hat{b}_{\vec{k}}^{\dagger}\hat{c}_{\vec{k}}\right].
\end{equation}
The Hamiltonian thus has five bare parameters, namely the bare resonance positions $B_1$ and $B_2$, and the coupling constants $\Lambda_1$, $\Lambda_2$ and $\Lambda_{12}$.
{\yy
Since background scattering in the open channel is not included in the model we expect it to work for systems with vanishingly small background scattering length $a_{bg}$.}

\subsection{Two-Body Observables: Scattering Length, Effective Range and Binding Energy}

To describe the two-body sector we solve the Schr{\"o}dinger equation $(\hat{H}-E)|\psi_{2B}\rangle=0$ with the most general two-body Ansatz in the center-of-mass frame: 
\begin{equation}
|\psi_{2B}\rangle=\gamma\hat{c}_{\vec{k}=0}^{\dagger}|0\rangle+\beta\hat{b}_{\vec{k}=0}^{\dagger}|0\rangle+\int\frac{d^{3}k}{\left(2\pi\right)^{3}}\alpha_{\vec{k}}\hat{a}_{\vec{k}}^{\dagger}\hat{a}_{-\vec{k}}^{\dagger}|0\rangle.
\label{eq:3-channel:two-body wave function}
\end{equation}
In the following, all quantities are renormalized with respect to a naturally arising momentum cut-off $k_c$ and its associated energy $E_c=\hbar^2k_c^2/m$.
For clarity, a dimensionful quantity $x$ is denoted $\tilde{x}$ after it is renormalized and dimensionless, for example, the renormalized scattering length $a$ (dimensionful) is denoted $\tilde{a}$ (dimensionless) and they are related via $\tilde{a}=k_ca$.
However, dimensions of magnetic field are not renormalized.
The molecular magnetic moment $\mu_i$ (dimensions of energy per magnetic field: J/G) is renormalized to $\tilde{\mu}_i=\mu_i/E_c$ which has dimensions of 1/G.

For positive energy $E=\hbar^2k_0^2/m>0$ the Schr{\"o}dinger equation leads to two coupled equations for the molecular amplitudes $\beta$ and $\gamma$:
\begin{subequations}
\begin{multline}
\left(\tilde{\mu}_{1}\left(B_{1}-B\right)-\tilde{k}_{0}^{2}\right)\tilde{\beta}+2\tilde{\Lambda}_{1}+\tilde{\Lambda}_{12}\tilde{\gamma} \\
-\frac{\tilde{\Lambda}_{1}}{\pi^{2}}\left(1-\frac{i\pi}{2}\tilde{k}_{0}\right)\left(\tilde{\Lambda}_{1}\tilde{\beta}+\tilde{\Lambda}_{2}\tilde{\gamma}\right)=0
\end{multline}
\begin{multline}
\left(\tilde{\mu}_{2}\left(B_{2}-B\right)-\tilde{k}_{0}^{2}\right)\tilde{\gamma}+2\tilde{\Lambda}_{2}+\tilde{\Lambda}_{12}\tilde{\beta} \\
-\frac{\tilde{\Lambda}_{2}}{\pi^{2}}\left(1-\frac{i\pi}{2}\tilde{k}_{0}\right)\left(\tilde{\Lambda}_{1}\tilde{\beta}+\tilde{\Lambda}_{2}\tilde{\gamma}\right)=0
\end{multline}
\label{eq:3-channel:equations for beta and gamma (normalized)}
\end{subequations}
with which the scattering amplitude
\begin{equation}
\tilde{f}_{k_0}=-\frac{\tilde{\Lambda}_{1}\tilde{\beta}+\tilde{\Lambda}_{2}\tilde{\gamma}}{4\pi}
\label{eq:3-channel:equation for scattering amplitude (normalized)}
\end{equation}
is computed.
The resulting expression is expanded to second order in $\tilde{k}_0$ and compared to the effective range expansion $\tilde{f}_{k_0}^{-1}=-\tilde{a}^{-1}-i\tilde{k}_0+\tilde{r}_e\tilde{k}_0^2/2$ to find $\tilde{a}$ and $\tilde{r}_e$.

If instead of the scattering states ($E>0$) we search for a bound state solution $E=-\hbar^{2}\lambda_D^{2}/m<0$ ($\lambda_D>0$) the following equations are obtained for the binding wave number of the dimer $\tilde{\lambda}_D$ and the ratio $\chi=\tilde{\beta}/\tilde{\gamma}$:
\begin{subequations}
\begin{multline}
\left(\tilde{\mu}_{1}\left(B_{1}-B\right)+\tilde{\lambda}_D^{2}\right)\chi+\tilde{\Lambda}_{12} \\
-\frac{\tilde{\Lambda}_{1}}{\pi^{2}}\left(1-\tilde{\lambda}_D\frac{\pi}{2}\right)\left(\tilde{\Lambda}_{1}\chi+\tilde{\Lambda}_{2}\right)=0
\end{multline}
\begin{multline}
\left(\tilde{\mu}_{2}\left(B_{2}-B\right)+\tilde{\lambda}_D^{2}\right)+\tilde{\Lambda}_{12}\chi \\
-\frac{\tilde{\Lambda}_{2}}{\pi^{2}}\left(1-\tilde{\lambda}_D\frac{\pi}{2}\right)\left(\tilde{\Lambda}_{1}\chi+\tilde{\Lambda}_{2}\right)=0
\end{multline}
\label{eq:3-channel:equations for lambda_D (normalized)}
\end{subequations}
By eliminating $\chi$, the two coupled Eqs.~(\ref{eq:3-channel:equations for lambda_D (normalized)}) can be written as a fourth order polynomial equation in $\tilde{\lambda}_D$.
Dependent on the value of $B$ it has two, one or no positive solutions.
The values of $B$ at which the number of solutions changes coincides with the resonance positions $a\rightarrow\pm\infty$ which we denote $B_i^{(\text{res})}$.

Details on the derivation of Eqs.~(\ref{eq:3-channel:equations for beta and gamma (normalized)}),~(\ref{eq:3-channel:equation for scattering amplitude (normalized)}) and~(\ref{eq:3-channel:equations for lambda_D (normalized)}) can be found in Appendix~\ref{app:3-channel:2-body}.

\subsection{Relating the Bare Parameters to Observables}
\label{sec:3-channel:bare to observable parameters}

Eliminating the amplitudes $\tilde{\beta}$ and $\tilde{\gamma}$ from Eqs.~(\ref{eq:3-channel:equations for beta and gamma (normalized)}) with $\tilde{k}_0=0$ one finds an analytic expression for $\tilde{a}=-\tilde{f}_{k_0=0}=(\tilde{\Lambda}_{1}\tilde{\beta}+\tilde{\Lambda}_{2}\tilde{\gamma})|_{k_0=0}/4\pi$ which can be parametrized as:
\begin{equation}
\tilde{a}(B)=\frac{\tilde{\Delta}_1}{B^{(\text{res})}_1-B}+\frac{\tilde{\Delta}_2}{B^{(\text{res})}_2-B},
\label{eq:3-channel:Feshbach resonance formula}
\end{equation}
where the resonance widths $\tilde{\Delta}_i$ and the positions $B^{(\text{res})}_i$ (for $i=1,2$) are observable parameters~\cite{DeltaVSDeltaB}.
This parametrization is also obtained in the context of multi-channel quantum defect theory~\cite{Jachymski13} {\yy (see Appendix~\ref{app:3-channel:compare to Jachymski})} and, therefore, generic.
Analytic expressions relating the four observable and five bare parameters are given in Appendix~\ref{app:3-channel:2-body analytic expressions}.
As expected, the observable parameters do not depend on the absolute position of the bare resonances but only on the difference $B_1-B_2$, except for $B^{(\text{res})}_i$ which also depend additively on the mean $(B_1+B_2)/2$ for positioning.
Because there is one more bare parameter than there are observable parameters there is an apparent redundancy in the model.
Indeed, keeping the observable parameters fixed, one can, for example, find a set of parameters $(\tilde{\Lambda}_1,\tilde{\Lambda}_2,B_1,B_2)$ for a variety of $\tilde{\Lambda}_{12}$ values without altering the scattering amplitude, the dimer binding energy or the trimer binding energy.
This is further discussed in Sec.~\ref{sec:Lambda_bb} below.

\subsection{Three-Body Sector: Efimov Trimers}

Here the main result of this paper, the equation for the trimer binding energy in the three-channel model, is stated.
Details of the derivation can be found in Appendix~\ref{app:3-channel:3-body}.
In short, the trimer binding energy $E_T=-\hbar^2\lambda_T^2/m$, with $\lambda_T>\max(0,\lambda_D)$, is the eigen value associated with the three-body wave function:
\begin{multline}
|\psi_{3B}\rangle=\int\frac{d^{3}k}{\left(2\pi\right)^{3}}\gamma_{\vec{k}}\hat{c}_{\vec{k}}^{\dagger}\hat{a}_{-\vec{k}}^{\dagger}|0\rangle+\int\frac{d^{3}k}{\left(2\pi\right)^{3}}\beta_{\vec{k}}\hat{b}_{\vec{k}}^{\dagger}\hat{a}_{-\vec{k}}^{\dagger}|0\rangle \\
+\int\frac{d^{3}k}{\left(2\pi\right)^{3}}\int\frac{d^{3}q}{\left(2\pi\right)^{3}}\alpha_{\vec{k},\vec{q}}\hat{a}_{\vec{q}+\frac{\vec{k}}{2}}^{\dagger}\hat{a}_{-\vec{q}+\frac{\vec{k}}{2}}^{\dagger}\hat{a}_{-\vec{k}}^{\dagger}|0\rangle.
\label{eq:3-channel:three-body wave function}
\end{multline}
Hence one must solve the Schr{\"o}dinger equation $(\hat{H}-E_T)|\psi_{3B}\rangle=0$ to arrive at a closed equation for $\lambda_T$.
The condition $\lambda_T>\max(0,\lambda_D)$ implies that only trimers associated with the higher resonance are properly determined by the following equations.
In between the two dimers, where Efimov trimers associated with the lower resonance are expected (see Fig.~\ref{fig:illustration}), a solution for any value of $\lambda_T$ exists due to the dimer-atom continuum.
It is not possible to distinguish between the dimer-atom and the trimer state since both are of the form~(\ref{eq:3-channel:three-body wave function})~\cite{dimerAtomScatteringResonances}.

Direct substitution of $|\psi_{3B}\rangle$ into $(\hat{H}-E_T)|\psi_{3B}\rangle=0$ leads to three coupled integral equations which are reduced to two by eliminating the free particle amplitude $\alpha_{\vec{k},\vec{q}}$.
It is then convenient to write the two remaining three-body scattering amplitudes as a vector $\psi(k)=(\beta_k,\gamma_k)^T$ and put the coefficients in a $2\times2$ matrix $\mathcal{M}_{\lambda_T}\left(k,q\right)$ that depends on $\lambda_T$.
The Schr{\"o}dinger equation thus takes the form $\int_0^\infty dq \mathcal{M}_{\lambda_T}(k,q)\psi(q)=0$ and a non-trivial solution is obtained for $\det\mathcal{M}_{\lambda_T}(k,q)=0$.
After renormalizing with respect to the momentum cut-off and using the practical substitution $k=(2/\sqrt{3})\lambda_T\sinh\xi$, the Schr{\"o}dinger equation can be written as:
\begin{equation}
\int_{-\infty}^{\infty} d\xi \mathcal{M}_{\lambda_T}\left(\xi,\xi^\prime\right)\psi\left(\xi^\prime\right)=0.
\label{eq:3-channel:int d xi M(xi,xi') psi(xi') = 0}
\end{equation}
The lower integration limit was extended to $-\infty$ by demanding that both $\tilde{\beta}_\xi$ and $\tilde{\gamma}_\xi$ be odd functions of $\xi$.
The vector $\psi(\xi)$ is now defined as $\psi(\xi)=(\tilde{\beta}_\xi,\tilde{\gamma}_\xi)^T$ and the matrix elements are:
\begin{multline}
\left(\mathcal{M}_{\lambda_T}\right)_{ij}=\bigg[\left(f_i\left(\xi^\prime\right)-h\left(\xi^\prime\right)\right)\delta_{ij}+h\left(\xi^\prime\right) \\
-\tilde{\Lambda}_i\tilde{\Lambda}_j g\left(\xi^\prime\right)\bigg]\delta\left(\xi-\xi^\prime\right)
-\tilde{\Lambda}_i\tilde{\Lambda}_j L\left(\xi,\xi^\prime\right),
\end{multline}
where we have defined:
\begin{subequations}
\begin{equation}
f_{i}\left(\xi\right)=\tilde{\lambda}_T\cosh\xi+\frac{\tilde{\mu}_{i}}{\tilde{\lambda}_T\cosh\xi}\left(B_{i}-B\right),
\end{equation}
\begin{equation}
g\left(\xi\right)=\frac{1}{\pi^{2}}\left(\frac{1}{\tilde{\lambda}_T\cosh\xi}-\frac{\pi}{2}\right),
\end{equation}
\begin{equation}
h\left(\xi\right)=\frac{\tilde{\Lambda}_{12}}{\tilde{\lambda}_T\cosh\xi},
\end{equation}
\begin{equation}
L\left(\xi,\xi^{\prime}\right)=\frac{2}{\sqrt{3}\pi^{2}}\ln\left(\frac{e^{2\left(\xi-\xi^{\prime}\right)}+e^{\xi-\xi^{\prime}}+1}{e^{2\left(\xi-\xi^{\prime}\right)}-e^{\xi-\xi^{\prime}}+1}\right).
\end{equation}
\end{subequations}
The requirement of a vanishing determinant:
\begin{equation}
\det \mathcal{M}_{\lambda_T}\left(\xi,\xi^\prime\right)=0,
\label{eq:3-channel:equantion for lambda_T, determinant = 0}
\end{equation}
is a closed equation for $\lambda_T$.
Depending on the magnetic field there are many values $\lambda_T=\lambda_T^{(\text{sol})}$ for which Eq.~(\ref{eq:3-channel:equantion for lambda_T, determinant = 0}) is satisfied.
To single out the physical solutions corresponding to three-body bound sates one must compute the zero-eigenvalue eigenfunction $\psi(\xi)$ of $\mathcal{M}_{\lambda_T^{(\text{sol})}}$ in accordance with Eq.~(\ref{eq:3-channel:int d xi M(xi,xi') psi(xi') = 0}) and determine $\tilde{\beta}_{\xi}$ and $\tilde{\gamma}_{\xi}$.
The mathematical solution $\lambda_T^{(\text{sol})}$ is physically relevant only if both are odd functions of $\xi$.
In addition, the number of nodes in $\tilde{\beta}_{\xi}$ and $\tilde{\gamma}_{\xi}$ allows the assignment of $\lambda_T^{(\text{sol})}$ to the ground or an excited Efimov state (see Sec.~\ref{sec:eigen functions}).

\section{Application to Model Systems}
\label{sec:model systems}

\subsection{Definitions}

\begin{table}[b]
\centering
\begin{tabular}{c | c c c}
\hline\hline
type & NB & NN & BN \\
\hline
$\Delta_1/a_0$ (G) & $150$ & $150$ & $1000$ \\
$\Delta_2/a_0$ (G) & $1000$ & $150$ & $150$ \\
\hline
$B_1-B_2^{(\text{res})}$ (G) & $-39.3857$ & $-23.7856$ & $-54.0702$ \\
$B_2-B_2^{(\text{res})}$ (G) & $-17.22$ & $-5.76371$ & $-2.53547$ \\
$\tilde{\Lambda}_1$ & $2.02991$ & $0.776438$ & $2.29268$ \\
$\tilde{\Lambda}_2$ & $1.21692$ & $0.926494$ & $0.587429$ \\
$\tilde{\Lambda}_{12}$ & $0.1$ & $0.1$ & $0.1$  \\
\hline\hline
\end{tabular}
\caption{\label{tb:model systems parameters}
Parameters of the three model systems.}
\end{table}

\begin{figure*}
\centering
\includegraphics[width=1.\linewidth]{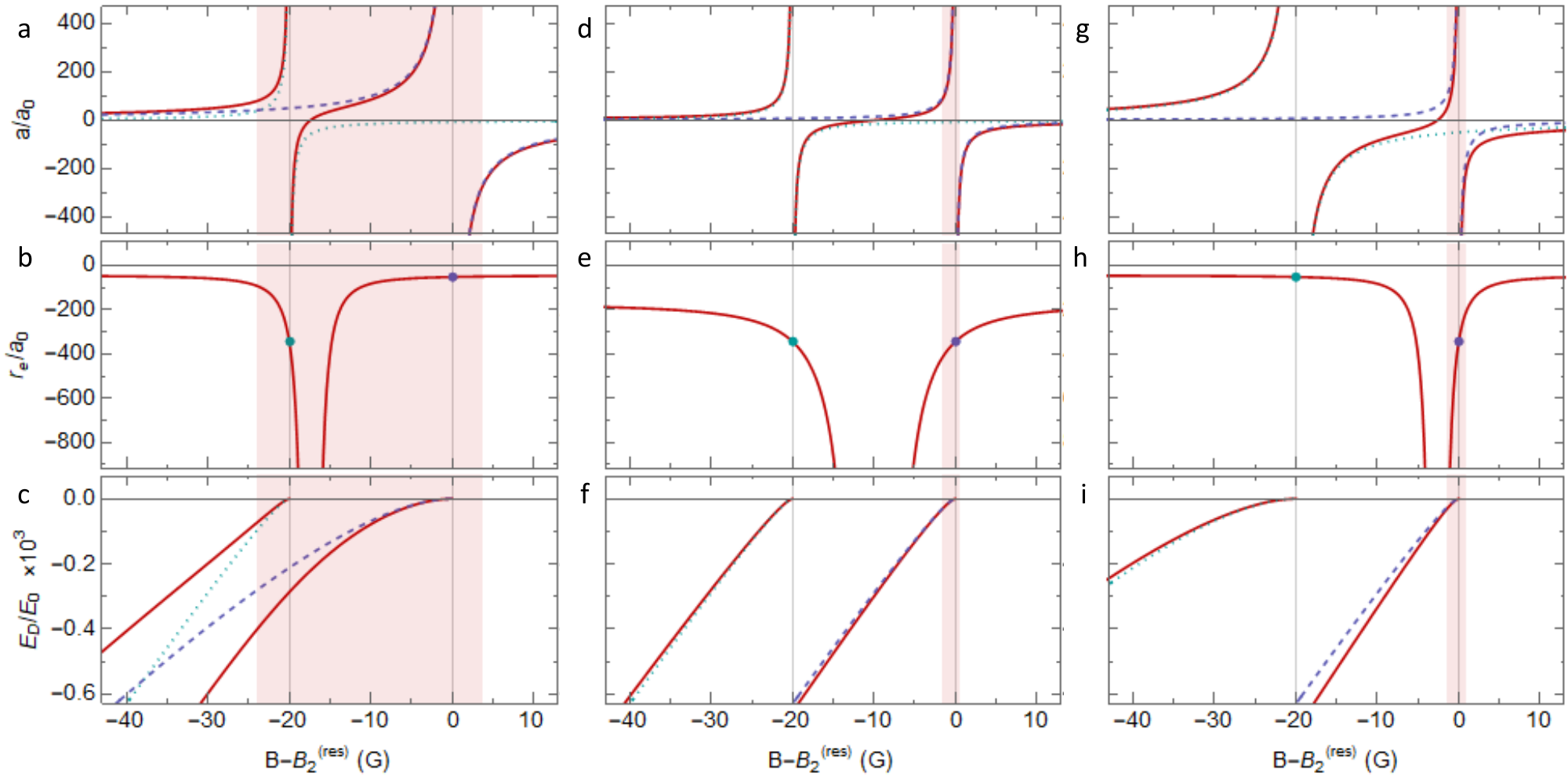}
\caption{\label{fig:two-body}
Two-body sector.
The scattering length (a,d,g), effective range (b,e,h) and dimer binding energy (c,f,i) for the NB (a,b,c), NN (d,e,f) and BN (g,h,i) scenarios are shown.
The three-channel model (solid curves) is compared to the two-channel model applied to the lower (dotted) and higher (dashed) resonance.
In (b,e,h) the two-channel value of the effective range is represented by a point because of its lack of $B$-dependence.
The grey vertical lines indicate the resonance positions $B_1^{(\text{res})}$ and $B_2^{(\text{res})}$, and the shaded region shows the extent of the ground state Efimov trimer $B_\star^{(0)}<B<B_-^{(0)}$ associated with $B_2^{(\text{res})}$.
}
\end{figure*}

\begin{figure*}
\centering
\includegraphics[width=1.\linewidth]{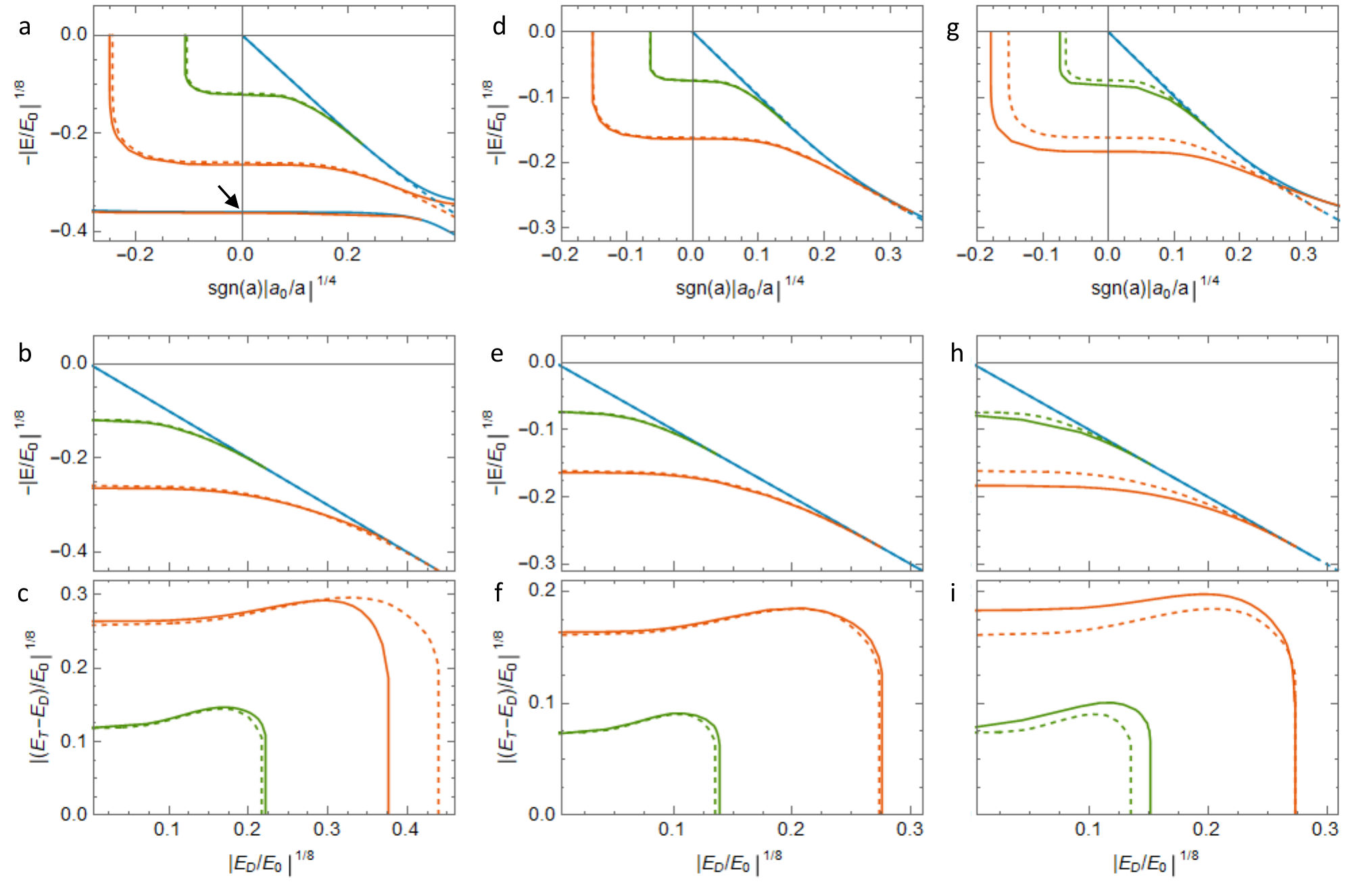}
\caption{\label{fig:three-body}
Three-body sector.
The dimer (blue), ground state trimer (orange) and excited state trimer (green) are plotted as a function of the inverse scattering length (a,d,g) and, for $a>0$, the dimer binding energy (b,e,h).
The difference between the trimer and the dimer is shown in (c,f,i).
As in Fig.~\ref{fig:two-body} the three columns correspond to the NB (a,b,c), NN (d,e,f) and BN (g,h,i) scenario.
The three-channel model (solid curves) is compared to the two-channel model (dashed).
{\yy
The arrow in (a) indicates where the lower resonance $B_1^{(\text{res})}$ is crossed.
}
}
\end{figure*}

{\yy
To illustrate the three-channel model we choose a model atom whose molecular bound states are pure spin singlets and consider high magnetic fields such that the Zeeman shift is linear to a good approximation.
The value of the magnetic moment is thus $\mu_1=\mu_2=-2\mu_B$, where $\mu_B=1.4$ MHz/G is the Bohr magneton.}
The momentum cut-off is somewhat arbitrarily fixed to $k_c=0.05/a_0$ but as discussed in Sec.~\ref{sec:Lambda_bb}, the results are indifferent to variations in $k_c$.
All lengths are calculated in units of the Bohr radius $a_0$ and all energies in units of $E_0=\hbar^2/ma_0^2$, where $m$ is the atomic mass.
Three scenarios are considered: $\Delta_1\ll\Delta_2$ (denoted NB), $\Delta_1=\Delta_2$ (NN) and $\Delta_1\gg\Delta_2$ (BN) -- see Table~\ref{tb:model systems parameters}.
The distance between the two resonances is identical in all three scenarios, we choose $B_2^{(\text{res})}-B_1^{(\text{res})}=20$ G, so the only difference between the models is the width.
{\yy In the NN scenario, as will be shown, $B_2^{(\text{res})}-B_1^{(\text{res})}$ is too large for the two resonances to be considered overlapping.
We therefore expect the results of the three-channel model to be in good agreement with those of the two-channel model, i.e. the additional channel has no influence on the two- and three-channel observables.
{\lk This system is used as a sanity check for our three-channel model.}
In the NB (BN) scenario the higher (lower) resonance is broadened to make them overlapping.
(Alternatively one could keep $\Delta_1=\Delta_2$ constant and decrease $B_2^{(\text{res})}-B_1^{(\text{res})}$ to generate overlap.)}
Because the three-channel model allows determination of the higher resonance trimers only, both the case NB and BN are considered.
In each scenario we ask the question: How does the resonance at $B_1^{(\text{res})}$ influence the Efimov spectrum around $B_2^{(\text{res})}$?

\subsection{Two-Body Sector}

From the analytic equations in Appendix~\ref{app:3-channel:2-body analytic expressions} we find the bare parameters for each model (Table~\ref{tb:model systems parameters}).
Here, because of the redundancy in number of bare parameters, we fix $\tilde{\Lambda}_{12}=0.1$.
Other options and their consequences are discussed in Secs.~\ref{sec:Lambda_bb} and~\ref{sec:eigen functions}.

We use the two-body equations to compute the scattering lengths, effective ranges and dimer binding energies of the three scenarios.
The results are compared to an individual treatment of the resonances with the two-channel model (Fig.~\ref{fig:two-body}).
As expected, the two-channel model is a good approximation only in the direct vicinity of the resonance.
The three-channel model introduces three important additions.
(1) The scattering length is forced to cross $a=0$ in between the two resonances, close to the narrower one.
According to the two-channel model this never happens (for zero background scattering length).
(2) While in the two-channel model the effective range is constant across the Feshbach resonance, it develops a magnetic field dependence in the three-channel model.
In particular, at $a=0$, $r_e\rightarrow-\infty$.
(3) Finally, unlike the two independent dimer energy levels arising from an individual treatment of the two resonances with the two-channel model, level repulsion naturally arises in the three-channel model.
We stress at this point that the level repulsion is not due to $\tilde{\Lambda}_{12}$ but is intrinsic to the model and also happens for $\tilde{\Lambda}_{12}=0$.
{\yy
The physical origin of the repulsion in this case is the second-order coupling through the continuum via $\hat{H}_1$ and $\hat{H}_2$.
Changing $\tilde{\Lambda}_{12}$ (and accordingly also $\tilde{\Lambda}_{1}$ and $\tilde{\Lambda}_{2}$, see Sec.~\ref{sec:Lambda_bb}) tunes the relative strength of the two.
}

The NN scenario in Fig.~\ref{fig:two-body}(d,e,f) is hardly affected by the additional channel.
In particular the two dimers are nearly identical in both treatments.
{\yy As mentioned above,}
this is the consequence of $B_2^{(\text{res})}-B_1^{(\text{res})}$ being large compared to $\Delta_1=\Delta_2$ and the resonances cannot be considered properly overlapping.
{\yy
The three-channel model thus reproduces the results of the two-channel model in the limit of non-overlapping resonances.
}
{\yy
Nevertheless, all three additions of the three-channel model, however small, are appreciable.
}

In the NB scenario on the other hand extensive repulsion of the two dimers is visible [Fig.~\ref{fig:two-body}(a,b,c)].
The two-channel model dimers are not coupled and intersect each other at a binding energy of $-0.57\cdot10^{-3}E_0$.
The mutual coupling introduced in the three-channel model leads to an avoided crossing, strongly altering their functional form.
{\yy
This is a stark contrast to the two-channel model already at the two-body level.
}

Finally, to lesser extent, this repulsion can also be seen in the BN scenario; Fig.~\ref{fig:two-body}(g,h,i).
The effect is much weaker because the two-channel model dimers don't cross.
In addition, a scattering length zero-crossing and its associated effective range divergence are introduced relatively close to the narrow resonance due to the neighboring broader resonance.

{\yy
We note that the scattering length zero-crossing could also be obtained in the two-channel model by using a non-zero background scattering length to account for the lower resonance.
Although the magnetic field regime for which the two- and three-channel scattering length agree would be extended, it would remain limited to the vicinity of the resonance.
The use of a non-zero background scattering length also raises the question of how to define it.
Does one prefer a larger regime of agreement or a perfect overlap at the zero-crossing?
Using the lower resonance explicitly avoids these questions, automatically takes into account the background generated by the nearby resonance and, most importantly, makes {\lk it} magnetic field dependent.

A more general model than we are presenting here would include both resonances and, in addition, a background scattering which arises from scattering in the open channel -- not another closed channel.
}

\subsection{Three-Body Sector}

Here we solve the three-body equations for the three scenarios.
In each case we compute the ground and first excited Efimov state around $B_2^{(\text{res})}$.
They are plotted in Fig.~\ref{fig:three-body} as a function of the inverse scattering length and as a function of the dimer binding energy.
The three-channel model is compared to the solution obtained from an isolated resonance according to the two-channel model.
We denote the scattering length value at which the $n$-th trimer (starting from the ground state $n=0$) crosses the free-atom continuum by $a_-^{(n)}$ and the value at which it merges with the dimer-atom continuum by $a_\star^{(n)}$.
The corresponding magnetic field values are denoted $B_-^{(n)}$ and $B_\star^{(n)}$, respectively.
To put the extent of the Efimov spectrum in context, Fig.~\ref{fig:two-body} highlights the region $B_\star^{(0)}<B<B_-^{(0)}$ for the three scenarios.

As in the two-body sector, the NN scenario is hardly affected by the additional molecule, demonstrating {\yy again} that the three-channel model reduces to the two-channel model for large $B_2^{(\text{res})}-B_1^{(\text{res})}$ and small $\Delta_{1,2}$ [Fig.~\ref{fig:three-body}(d,e,f)].
The three-channel trimers almost overlap with the two-channel trimers, although they are pushed to slightly deeper binding energies.

For smaller $B_2^{(\text{res})}-B_1^{(\text{res})}$ the overlap grows and the deepening effect increases.
As a real world example one may consider the $bb$ channel of $^{39}$K which features two very close resonances ($B_2^{(\text{res})}-B_1^{(\text{res})}=6.8$ G) of comparable widths.
Indeed the $\Delta$-parameters are within $10\%-15\%$ of $\Delta_1/a_0=\Delta_2/a_0=150$ G while $B_2^{(\text{res})}-B_1^{(\text{res})}$ is a factor of $\sim3$ smaller than in our NN scenario.
Experiments have shown deviations from the prediction of the two-channel model~\cite{Roy13}.
In fact, $a_-^{(0)}$ was found to be at a lower scattering length value than predicted, in agreement with the general trend pointed out by the three-channel model.
{\yy
Using coupled-channels values for the experimentally relevant parameters~\cite{JuliennePrivate} the three-channel model predicts $a_-^{(0)}\approx-6030a_0$.
This corresponds to a shift of $\sim8\%$ with respect to the the two-channel model value $a_-^{(0)}\approx-6550a_0$.
The {\lk reported} experimental value is $a_-^{(0)}\approx-1000a_0$.
A quantitative comparison to the experiment is, {\lk however}, inconvenient since the measurements, as pointed out in Ref.~\cite{Roy13}, are accompanied by large uncertainties {\lk which arise mainly from the fact that the functional form of the experimental results disagree with the known theoretical models.
Thus, the level of understanding of these results have yet {\yy to} reach a level which would allow a meaningful comparison with the three-channel model.
}
}

In the NB scenario, the most striking difference is that the ground state trimer extends from $B_-^{(0)}>B_2^{(\text{res})}$ to $B_\star^{(0)}<B_1^{(\text{res})}$ [see shaded region in Fig.~\ref{fig:two-body}(a,b,c)], i.e. it merges with the atom-dimer continuum after passing through the scattering length zero-crossing and {\lk the pole of} the lower resonance.
{\yy
This is manifested by $E_D$ and $E_T^{(0)}$ exiting the plot in Fig.~\ref{fig:three-body}(a) through $1/a\rightarrow\infty$, reentering from $1/a\rightarrow-\infty$, crossing $1/a=0$ again due to the lower resonance [see arrow in Fig.~\ref{fig:three-body}(a)] to finally merge in the $a>0$ region.
Note that the Efimov trimer remains bound even though the scattering length vanishes.}
{\yy
The two-channel model can't possibly capture this effect due to the absence of the second molecular channel, even if one would include a non-zero background scattering length.
{\lk In fact, the background scattering length induced by the second resonance obviously diverges at {\it its} pole.
This strong magnetic field dependence makes the extension of the two-channel model by any finite (constant) background scattering length ineffective.}
}
In addition, although the three-channel trimer is slightly deeper bound than the two-channel trimer for most of the spectrum, as they approach $a_\star^{(0)}$ the two cross and the three-channel trimer merges at a larger scattering length value [Fig.~\ref{fig:three-body}(b,c)].
The effect on the excited trimer is very similar to the NN scenario.

A real world example for the NB scenario is the $bb$ channel in $^7$Li.
It features two resonances with $R^\star_1=722a_0$ ($s_{res,1}=0.0411$), $R^\star_2=60a_0$ ($s_{res,2}=0.493$) and $B_2^{(\text{res})}-B_1^{(\text{res})}=48.4$ G~\cite{Gross11,Jachymski13,Julienne14}.
Indeed, the three-channel model is very good in the two-body sector.
It reproduces the scattering length, effective range and dimer binding energies better than the individual two-channel model treatments (see Appendix~\ref{app:Li-7}).
However, the three-body sector is unexpectedly dominated by finite-range corrections despite both resonances having $s_{res}<1$~\cite{Machtey12}.
The reason for this behavior is currently unknown and considered an open question in few-body physics~\cite{Schmidt12,Langmack18}.
This puzzle is beyond the reach of our simplified model which neglects all van der Waals physics.

The trimer in the BN scenario is pushed to lower energies by an appreciable amount as shown in Fig.~\ref{fig:three-body}(g).
In particular, the Efimov resonance at $a_-^{(0)}$, is shifted from $R_\star/a_-^{(0)}=-0.092$ in the two-channel model~\cite{Nishida12} to $R_\star/a_-^{(0)}=-0.174$.
This factor of $\sim2$ reduction in the absolute value of $a_-^{(0)}$ works in favor of the experimental demonstration of Efimov physics in the vicinity of narrow resonances, as it relaxes severe magnetic field stability requirements necessary for experimental exploration of narrow Feshbach resonances~\cite{Johansen17}. 
On the $a>0$ side one notes that $a_\star^{(0)}$ is unaltered but the excited state $a_\star^{(1)}$ is [see Fig.~\ref{fig:three-body}(h,i)].
In addition, away from $a_\star^{(n)}$ ($n=0,1$) both states are pushed to deeper binding energies and reach maximal deviation from the two-channel model at the resonance position where $E_D\rightarrow0$.
{\lk We note that in this particular scenario 
{\yy a two-channel model which includes a non-zero background scattering length would be in better agreement with the three-channel results.}
However, we expect the three-channel model to be superior, because it naturally includes the magnetic field dependence of the background scattering length.
It also removes the unavoidable ambiguity of choosing a certain background scattering length in the improved two-channel model.}

For the BN scenario we mention the $aa$ channel of the heteronuclear $^6$Li-Cs system as a real world example.
Its two resonances have $s_{res,1}=0.66$ and $s_{res,2}=0.05$ and their distance is $B_2^{(\text{res})}-B_1^{(\text{res})}=49.9$ G.
In fact, this is the only system to date, where Efimov resonances near a truly narrow Feshbach resonance (i.e. for a resonance with $s_{res}\ll1$ as opposed to $s_{res}\lesssim1$) were measured~\cite{Johansen17}.
The experimental value of $a_-^{(1)}$ associated with the $B_2^{(\text{res})}$ trimers was found to be significantly lower than predicted by {\lk two-channel} theory.
Quantitative analysis of this system with the three-channel model requires extension of the latter to the hetero-nuclear case.

\section{Discussion of the Free Bare Parameter}
\label{sec:Lambda_bb}

\begin{figure}
\centering
\includegraphics[width=1.\linewidth]{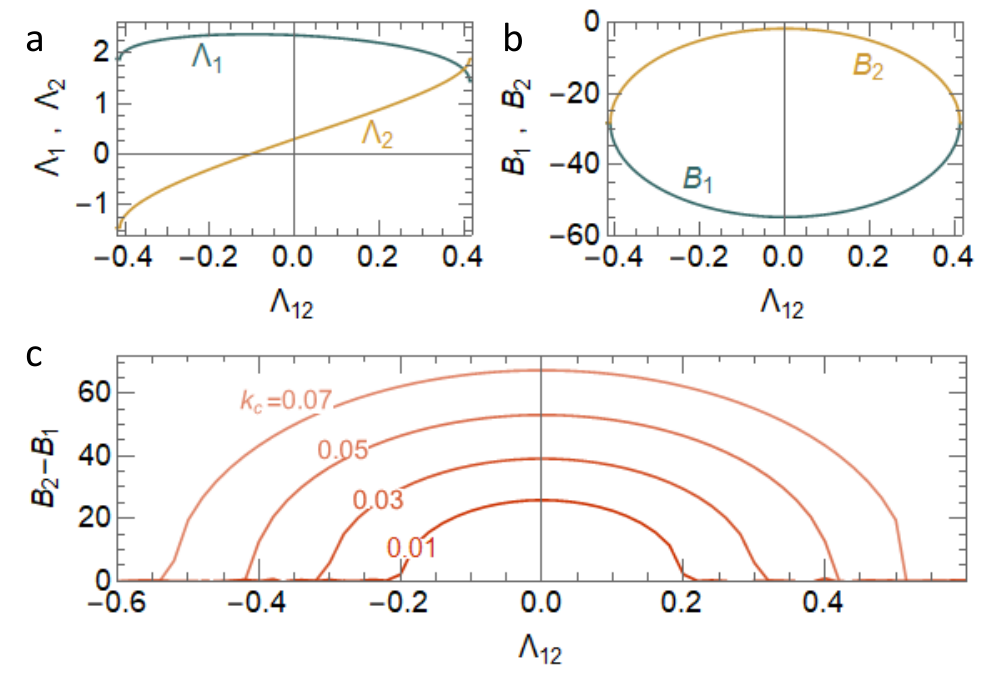}
\caption{\label{fig:Lambda_bb dependence}
Dependence of (a) $\tilde{\Lambda}_{1,2}$  and of (b) $B_{1,2}$ (with respect to $B_2^{(\text{res})}$) on $\tilde{\Lambda}_{12}$ for fixed $k_c$ and observable parameters.
(c) Dependence of $B_2-B_1$ on $\tilde{\Lambda}_{12}$ for various values of $k_c$ as indicated (in units of $1/a_0$) and fixed observable parameters.
}
\end{figure}

\begin{figure*}
\centering
\includegraphics[width=1.\linewidth]{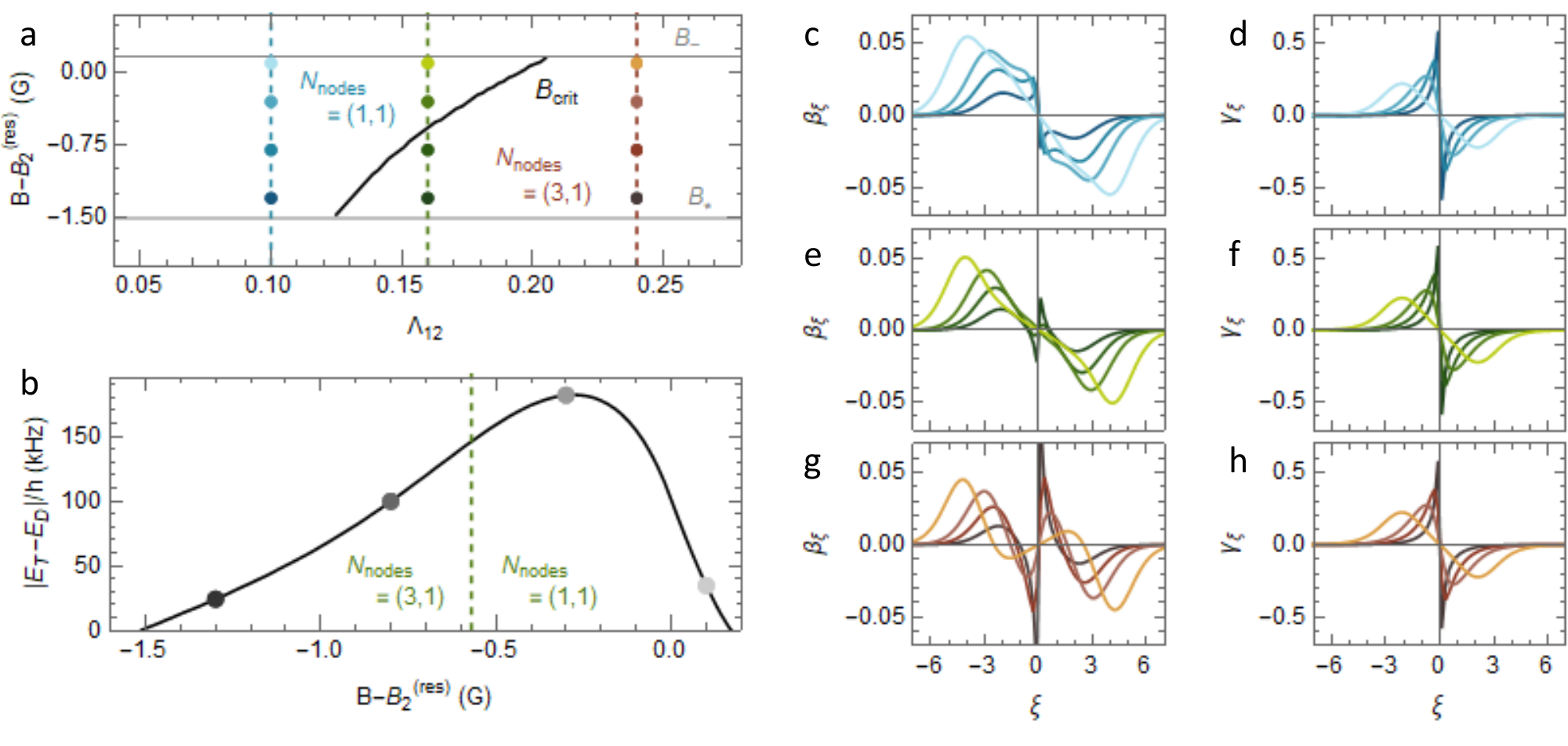}
\caption{\label{fig:eigen functions}
(a) {\yy Nodal pattern} of the BN scenario ground state.
Above and to the left of the black line, which is given by Eq.~(\ref{eq:3-channel:inequality for critical field (1+1) to (3+1) nodes}), $\tilde{\beta}_\xi$ has one node such that $N_{\text{nodes}}=\left(1,1\right)$.
Below and to the right $N_{\text{nodes}}=\left(3,1\right)$.
(b) Plot of the ground state Efimov energy with respect to the dimer-atom (for $B<B_2^{(\text{res})}$) and free-atom (for $B>B_2^{(\text{res})}$) continuum.
The dashed line indicates the transition from $N_{\text{nodes}}=\left(3,1\right)$ to $\left(1,1\right)$ in the case of $\tilde{\Lambda}_{12}=0.16$.
(c,e,g) Plot of $\tilde{\beta}_\xi$ and (d,f,h) of $\tilde{\gamma}_\xi$ for the three values of $\tilde{\Lambda}_{12}$ and the four values of $B$ indicated by the points in (a).
The points in (b) also indicate the values of $B$.
}
\end{figure*}

As mentioned in Sec.~\ref{sec:3-channel:bare to observable parameters} (see also Appendix~\ref{app:3-channel:2-body analytic expressions}) there are four observable parameters, namely $\{B_{1,2}^{(\text{res})},\Delta_{1,2}\}$, related to the five bare parameters $\{B_{1,2},\tilde{\Lambda}_{1,2,12}\}$ of the model.
Hence, one of the latter is free to choose.
We emphasize though, that as long as the observable parameters are fixed, and therefore constrain the bare parameters to change in a mutually dependent manner, all two- and three-body observables (such as scattering length and binding energies) remain the same.

For illustration, in what follows, we treat $\tilde{\Lambda}_{12}$ as the free parameter and fix the observable parameters to those of the BN scenario.
The $\tilde{\Lambda}_{12}$-dependence of the other four bare parameters is shown in Fig.~\ref{fig:Lambda_bb dependence}(a,b) for fixed cut-off $k_c=0.05/a_0$.
As $|\tilde{\Lambda}_{12}|$ increases the bare resonance position difference $B_2-B_1$ decreases towards $0$.
Beyond the point where $B_2-B_1=0$ no solution for the analytic equations of Appendix~\ref{app:3-channel:2-body analytic expressions} exists.
Therefore, there is a maximal value $\tilde{\Lambda}_{12}^{\text{max}}$ which the intermolecular coupling constant can assume.

Although $\tilde{\Lambda}_{12}$ may in principle be chosen anywhere in the range $-\tilde{\Lambda}_{12}^{\text{max}}<\tilde{\Lambda}_{12}<\tilde{\Lambda}_{12}^{\text{max}}$, additional physical inputs can constrain $\tilde{\Lambda}_{12}$.
For example, the absolute position of $B_{1,2}$ or their relative position $B_2-B_1$ introduces a fifth condition on the bare parameters.
Alternatively, one may wish to maintain $|\tilde{\Lambda}_1|>|\tilde{\Lambda}_2|$, as is the case in the two-channel model.
See, for example, the case of the $bb$-channel of $^7$Li considered in Appendix~\ref{app:Li-7}.

Fig.~\ref{fig:Lambda_bb dependence}(c) shows the relative position $B_2-B_1$ as a function of $\tilde{\Lambda}_{12}$ for various values of $k_c$.
The value of $\tilde{\Lambda}_{12}^{\text{max}}$ increases together with $k_c$ and so does the range of $B_{1,2}$.
Physically, $k_c$ should be chosen on the order of the inverse potential range, i.e. $k_c\sim1/r_{vdW}$.
Variations have no influence on the two- and three-body observables though.
Also here, additional physical inputs of a real system may further constrain the value of $k_c$.

\section{Trimer Eigen Functions}
\label{sec:eigen functions}

The matrix $\mathcal{M}_{\lambda_T^{(\text{sol})}}(\xi,\xi^\prime)$, with $\lambda_T^{(\text{sol})}$ a solution of Eq.~(\ref{eq:3-channel:equantion for lambda_T, determinant = 0}), has one vanishing eigenvalue in accordance with Eq.~(\ref{eq:3-channel:int d xi M(xi,xi') psi(xi') = 0}).
We compute the corresponding eigenfunction $\psi_0(\xi)$ satisfying $\int d\xi^\prime \mathcal{M}_{\lambda_T^{(\text{sol})}}(\xi,\xi^\prime)\psi_0(\xi^\prime)=0$ and extract the amplitudes of $\psi_0(\xi)=(\tilde{\beta}_\xi,\tilde{\gamma}_\xi)^T$.
Recall that the amplitudes that correspond to a physical bound state are odd functions of $\xi$ (odd number of nodes) which implies that they vanish at $\xi=0$.
Solutions of Eq.~(\ref{eq:3-channel:equantion for lambda_T, determinant = 0}) that lead to an even zero-eigenvalue eigenfunction are discarded on this basis.

\begin{figure}[b]
\centering
\includegraphics[width=0.7\linewidth]{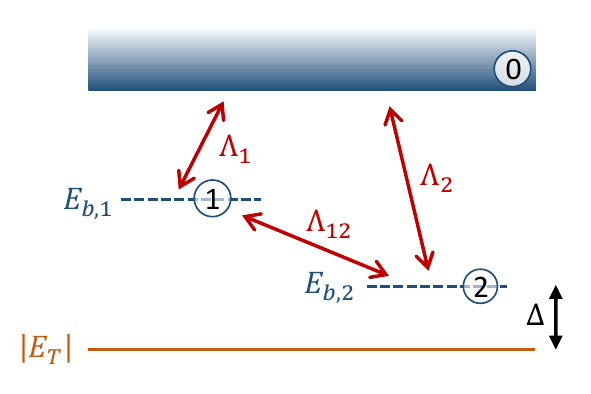}
\caption{\label{fig:direct VS indirect}
Analogy of the three-channel model to a three-level system.
The circled numbers are the quantum optics energy levels.
Here, the two-photon detuning is $\Delta=\left|E_T\right|+E_{b,2}$.
}
\end{figure}

For specificity the following discussion focuses on the ground state of the BN scenario, whose binding energy is shown in Fig.~\ref{fig:eigen functions}(b), but the {\lk conclusions are} general.
The amplitudes $\tilde{\beta}_\xi$ and $\tilde{\gamma}_\xi$ for the three values $\tilde{\Lambda}_{12}=0.1$, $0.16$ and $0.24$ are shown in Fig.~\ref{fig:eigen functions}(c,d) , (e,f) and (g,h) respectively.
While $\tilde{\gamma}_\xi$ is insusceptible and has a single node (at $\xi=0$), the form of $\tilde{\beta}_\xi$, whose amplitude is an order of magnitude smaller, is sensitive to changes in $\tilde{\Lambda}_{12}$.
Fig.~\ref{fig:eigen functions}(c,e,g) shows how the number of nodes changes from one to three as $\tilde{\Lambda}_{12}$ is increased.
For certain values of $\tilde{\Lambda}_{12}$, see Fig.~\ref{fig:eigen functions}(b,e), the number of nodes depends on the position within the spectrum.
There is a critical magnetic field value $B_{\text{crit}}$ above (below) which $\tilde{\beta}_\xi$ has one (three) nodes.
Moreover, $B_{\text{crit}}$ depends on $\tilde{\Lambda}_{12}$ and therefore gives rise to the {\yy nodal pattern} represented in Fig.~\ref{fig:eigen functions}(a).
For convenience we denote $N_\text{nodes}=(\text{number of nodes in }\tilde{\beta}_\xi,\text{number of nodes in }\tilde{\gamma}_\xi)$ such that $N_\text{nodes}=(1,1)$ to the left of the black curve and $N_\text{nodes}=(3,1)$ to the right.

The number of nodes in a wave function is indicative of the excitedness of the state.
For example, in the two-channel model (see Appendix~\ref{app:2-channel}), since only odd wave functions are allowed, the $n$-th state (starting at the ground state $n=0$) has $2n+1$ nodes.
In the three-channel model, $\tilde{\gamma}_\xi$, which is the dominant molecule-atom amplitude for the $B_2^{(\text{res})}$ trimers, follows this rule.
The secondary molecule-atom amplitude $\tilde{\beta}_\xi$ on the other hand may have $2n+1$ or $2(n+1)+1$ nodes, signifying it being in the $n$-th or $(n+1)$-th state.
{\yy As we will show, this is the result of two competing processes whose amplitudes are proportional to $\Lambda_1$ and $\Lambda_{12}\Lambda_2$.}
This could serve as an experimental indicator for the value of $\Lambda_{12}$.

To find an equation for the $\tilde{\Lambda}_{12}$-dependent $B_{\text{crit}}$ we consider the ratio $\chi_\xi=\tilde{\beta}_\xi/\tilde{\gamma}_\xi$.
At $\xi=0$, where both amplitudes have a node, $\chi_\xi$ remains finite.
It vanishes only if $\tilde{\beta}_\xi$ has a node while $\tilde{\gamma}_\xi$ does not and is thus indicative of the excess number of nodes in $\tilde{\beta}_\xi$.
Switching back to $k$ via $k=(2/\sqrt{3})\lambda_T\sinh\xi$, the following expression for $\chi_k$ can be readily derived (see Appendix~\ref{app:3-channel:3-body eigen function ratio}):
\begin{equation}
\chi_{k}=\frac{\tilde{\Lambda}_{1}\left[\frac{3}{4}\tilde{k}^{2}+\tilde{\lambda}_{T}^{2}+\tilde{\mu}_{2}\left(B_{2}-B\right)\right]-\tilde{\Lambda}_{2}\tilde{\Lambda}_{12}}{\tilde{\Lambda}_{2}\left[\frac{3}{4}\tilde{k}^{2}+\tilde{\lambda}_{T}^{2}+\tilde{\mu}_{1}\left(B_{1}-B\right)\right]-\tilde{\Lambda}_{1}\tilde{\Lambda}_{12}}.
\label{eq:3-channel:ratio chi_k in terms of bare parameters}
\end{equation}
This function features one zero-crossing for $\text{Re}\{k\}\geq0$ (and another for $\text{Re}\{k\}\leq0$) if
{\lk
\begin{equation}
\Lambda_1 < \frac{\Lambda_2\Lambda_{12}}{\left|E_T\right|+E_{b,2}}
\label{eq:3-channel:inequality for critical field (1+1) to (3+1) nodes}
\end{equation}
and none otherwise.
If Eq.~(\ref{eq:3-channel:inequality for critical field (1+1) to (3+1) nodes}) is satisfied the eigen functions correspond to $N_{\text{nodes}}=\left(3,1\right)$, if not, to $\left(1,1\right)$.
{\yy
To give meaning to the inequality we rearrange the channels as depicted in Fig.~\ref{fig:direct VS indirect} and draw the analogy to a three-level system in quantum optics.
While the left-hand-side of Eq.~(\ref{eq:3-channel:inequality for critical field (1+1) to (3+1) nodes}) is analogues to the Rabi frequency for the direct (one-photon) transition from level 0 to 1, the right-hand-side is the equivalent of the effective Rabi frequency for the indirect (two-photon) transition via level 2.}
Thus, the inequality states that the extra-node in $\beta_k$ (for $k>0$) is the result of the indirect coupling strength surpassing the direct one. 
}

For a given $\tilde{\Lambda}_{12}$, inequality~(\ref{eq:3-channel:inequality for critical field (1+1) to (3+1) nodes}) is solved for $B_{\text{crit}}$ and displayed in Fig.~\ref{fig:eigen functions}(a).
For the case $\tilde{\Lambda}_{12}=0.16$ we find $B_{\text{crit}}-B_2^{(\text{res})}=-0.57$ G as indicated in Fig.~\ref{fig:eigen functions}(b).

\section{Conclusions}
\label{sec:conclusions}
We have developed a simple three-channel theory of overlapping Feshbach resonances and show that the Efimov spectrum can be substantially altered in this scenario.
Experimental observations that are in disaccord with the isolated resonance theory can be revisited with the three-channel model (e.g. $^6$Li-Cs).
Moreover, given the demanding requirements for measuring Efimov resonances in the vicinity of a truly narrow resonance, our treatment allows identification of the favorable structure of Feshbach resonances.

The model can be generalized to fermionic systems~\cite{Castin06} and $\hat{H}_0$ can be extended to include background scattering as was done for the two-channel model~\cite{Werner09}.
In addition to three-body bound states one may analyze low-energy atom-dimer and three-atom scattering~\cite{Jona-Lasinio08,dimerAtomScatteringResonances} as well as four-body bound states associated with Efimov trimers~\cite{Castin10}.
Although cumbersome, one can speculate of an extension to $N$-channel theory for $N-1>2$ overlapping Feshbach resonances to describe possible few-body states in even more complex scenarios.

\section*{Acknowledgments}

We acknowledge fruitful discussions with F. Chevy, J. P. D'Incao and P. S. Julienne.
This research was supported in part by the Israel Science Foundation (Grant No. 1543/20) and by a grant from the United States-Israel Binational Science Foundation (BSF), Jerusalem, Israel, and the United States National Science Foundation.

\appendix

\section{Review of the Two-Channel Model}
\label{app:2-channel}

Here the two-channel model is reiterated in a slightly different approach than usual.
In particular, the momentum cut-off is used for renormalization purposes; see Eq.~(\ref{eq:2-channel:renormalization}) below.
Following the introduction of the Hamiltonian we show that the model provides analytic expressions for all two-body observables.
Thereafter an integral equation for the three-body bound states is derived.

\subsection{Two-Channel Hamiltonian}

We start from the Hamiltonian $\hat{H}=\hat{H}_{0}+\hat{H}_{int}$, where
\begin{equation}
\hat{H}_{0}=\int\frac{d^{3}k}{\left(2\pi\right)^{3}}\left[\frac{\hbar^{2}k^{2}}{2m}\hat{a}_{\vec{k}}^{\dagger}\hat{a}_{\vec{k}}+\left(E_{b}+\frac{\hbar^{2}k^{2}}{4m}\right)\hat{b}_{\vec{k}}^{\dagger}\hat{b}_{\vec{k}}\right]
\end{equation}
entails an open and a closed channel and
\begin{multline}
\hat{H}_{int}=\Lambda\int\frac{d^{3}k}{\left(2\pi\right)^{3}}\int\frac{d^{3}q}{\left(2\pi\right)^{3}} \\
\left[\hat{b}_{\vec{k}}^{\dagger}\hat{a}_{\vec{q}+\frac{\vec{k}}{2}}\hat{a}_{-\vec{q}+\frac{\vec{k}}{2}}+\hat{a}_{-\vec{q}+\frac{\vec{k}}{2}}^{\dagger}\hat{a}_{\vec{q}+\frac{\vec{k}}{2}}^{\dagger}\hat{b}_{\vec{k}}\right]
\end{multline}
couples them with coupling constant $\Lambda$.
Here, $\hat{a}_{\vec{k}}$ ($\hat{b}_{\vec{k}}$) annihilates an atom (a molecule) with momentum $\hbar\vec{k}$ and mass $m$ ($2m$) in the open (closed) channel and $\hat{a}_{\vec{k}}^\dagger$ ($\hat{b}_{\vec{k}}^\dagger$) is its Hermitian conjugate.
The bare molecular binding energy $E_b$ is assumed to be an affine function of an externally applied magnetic field $B$: $E_b=\mu(B_0-B)$, where $\mu$ is the molecules magnetic moment with respect to the open channel and $B_0$ is the bare resonance position.
The free parameters of the system are thus $\Lambda$ and $B_0$.

\subsection{Two-Body Observables: Scattering Length, Effective Range and Binding Energy}

To compute two-body observables the following wave function is used in the Schr{\"o}dinger equation $(\hat{H}-E)|\psi_{2B}\rangle$:
\begin{equation}
|\psi_{2B}\rangle=\beta\hat{b}_{\vec{k}=0}^{\dagger}|0\rangle+\int\frac{d^{3}k}{\left(2\pi\right)^{3}}\alpha_{\vec{k}}\hat{a}_{\vec{k}}^{\dagger}\hat{a}_{-\vec{k}}^{\dagger}|0\rangle.
\end{equation}
This center-of-mass superpostion of two free atoms and one bare molecule is the most general wave function for a two-body system and therefore a suitable Ansatz.
The two coupled equations
\begin{subequations}
\begin{equation}
\left(\frac{\hbar^{2}k^{2}}{m}-E\right)\alpha_{\vec{k}}+\Lambda\beta=0
\end{equation}
\begin{equation}
\left(E_{b}-E\right)\beta+2\Lambda\int\frac{d^{3}q}{\left(2\pi\right)^{3}}\alpha_{\vec{q}}=0
\end{equation}
\end{subequations}
are obtained.
Comparison of the open channel coefficient $\alpha_{\vec{k}}$, as obtained from the first equation for $E=\hbar^2 k_0^2/m>0$, with the scattering Green's function
\begin{equation}
G_{\text{scat}}\left(k,k_0\right)=\left(2\pi\right)^{3}\delta\left(\vec{k}-\vec{k}_{0}\right)+\frac{4\pi f_{k_{0}}}{k^{2}-k_{0}^{2}-i\eta}
\label{eq:Green's function for scattering problem}
\end{equation}
implies that the scattering amplitude is given via
\begin{equation}
f_{k_0}=-\frac{m\Lambda\beta}{4\pi\hbar^2}.
\label{eq:2-channel:scattering amplitude (not normalized)}
\end{equation}
Next, $\alpha_{\vec{k}}$ is plugged into the second equation and, exploiting the spherical symmetry of $s$-wave scattering $\alpha_{\vec{k}}=\alpha_k$, one finds an equation for the molecular amplitude $\beta$:
\begin{equation}
\left[\mu\left(B_{0}-B\right)-\frac{\hbar^{2}k_{0}^{2}}{m}\right]\beta+2\Lambda-\frac{m\Lambda^{2}}{\pi^{2}\hbar^{2}}\left(k_{c}-\frac{i\pi}{2}k_{0}\right)\beta=0.
\label{eq:2-channel:equation for beta (not normalized)}
\end{equation}
Here a high momentum cut-off $k_c$ was introduced during the computation of the radial integral to avoid a divergence.
If instead of $E>0$ one searches for a bound state solution $E=-\hbar^2\lambda_D^2/m<0$ (with $\lambda_D>0$) the same procedure leads to
\begin{equation}
\left[\mu\left(B_{0}-B\right)+\frac{\hbar^{2}\lambda_D^{2}}{m}\right]-\frac{m\Lambda^{2}}{\pi^{2}\hbar^{2}}\left(k_{c}-\lambda_D\frac{\pi}{2}\right)=0
\label{eq:2-channel:equation for lambda_D (not normalized)}
\end{equation}
where $\beta$ conveniently canceled.
In order to get rid off the momentum cut-off we renormalize the model parameters according to
\begin{equation}
\tilde{\Lambda}=\frac{\Lambda k_{c}^{3/2}}{E_{c}} ,\quad
\tilde{\mu}=\frac{\mu}{E_{c}} ,\quad
\tilde{\beta}=\beta k_{c}^{3/2}
\label{eq:2-channel:renormalization}
\end{equation}
where $E_c=\hbar^2k_c^2/m$.
In addition the scattering and binding wave number are renormalized as $\tilde{k}_0=k_0/k_c$ and $\tilde{\lambda}_D=\lambda_D/k_c$.
With this, Eqs.~(\ref{eq:2-channel:equation for beta (not normalized)}) and~(\ref{eq:2-channel:equation for lambda_D (not normalized)}) become
\begin{equation}
\left(\tilde{\mu}\left(B_{0}-B\right)-\tilde{k}_{0}^{2}\right)\tilde{\beta}+2\tilde{\Lambda}-\frac{\tilde{\Lambda}^{2}}{\pi^{2}}\left(1-\frac{i\pi}{2}\tilde{k}_{0}\right)\tilde{\beta}=0
\label{eq:2-channel:equation for beta (normalized)}
\end{equation}
and
\begin{equation}
\left(\tilde{\mu}\left(B_{0}-B\right)+\tilde{\lambda}_D^{2}\right)-\frac{\tilde{\Lambda}^{2}}{\pi^{2}}\left(1-\tilde{\lambda}_D\frac{\pi}{2}\right)=0,
\label{eq:2-channel:equation for lambda_D (normalized)}
\end{equation}
and the normalized scattering amplitude $\tilde{f}_{k_0}=k_c f_{k_0}$ is
\begin{equation}
\tilde{f}_{k_0}=\frac{\tilde{\Lambda}\tilde{\beta}}{4\pi}.
\end{equation}
By solving Eq.~(\ref{eq:2-channel:equation for beta (normalized)}) for $\tilde{\beta}$ and comparing $\tilde{f}_{\tilde{k}_0}$ to the known low-energy expansion
\begin{equation}
\frac{1}{\tilde{f}_{k_{0}}}=-\frac{1}{\tilde{a}}-i\tilde{k}_{0}+\frac{\tilde{r}_{e}\tilde{k}_{0}^{2}}{2},
\label{eq:scattering amplitude (normalized) effective range expansion}
\end{equation}
where $\tilde{a}=k_c a$ and $\tilde{r}_e=k_c r_e$ are the renormalized scattering length and effective range respectively, one finds
\begin{equation}
\tilde{a}=-\frac{1}{2\pi}\frac{\tilde{\Lambda}^{2}}{\tilde{\mu}\left(B_{0}-B\right)-\frac{\tilde{\Lambda}^{2}}{\pi^{2}}}
\end{equation}
and
\begin{equation}
\tilde{r}_{e}=-\frac{4\pi}{\tilde{\Lambda}^2}.
\label{eq:2-channel:effective range is proportional to 1/Lambda^2}
\end{equation}
We note that the effective range is field independent.
By denoting
\begin{equation}
B_{\text{res}}=B_{0}-\frac{\tilde{\Lambda}^{2}}{\tilde{\mu}\pi^{2}} ,\quad
\tilde{\Delta}=\frac{\tilde{\Lambda}^{2}}{2\pi\tilde{\mu}},
\end{equation}
the scattering length may be written in the familiar form
\begin{equation}
\tilde{a}=\frac{\tilde{\Delta}}{B-B_{\text{res}}}.
\label{eq:2-channel:Feshbach resonance formula}
\end{equation}
{\yy Here, $\Delta$ is defined with the opposite sign with respect to Eq.~(\ref{eq:3-channel:Feshbach resonance formula}).}
We note that this expression, which is faithfully reproduced, is far more general than the two-channel model~\cite{Jachymski13}.
The expression for $B_{\text{res}}$ demonstrates that the actual resonance position is shifted away from the bare resonance position $B_0$ by the coupling to the open channel.
Furthermore, the expression for $\tilde{\Delta}$ shows that a narrow resonance arises from weakly coupled channels as eluded to in the introduction.
From Eq.~(\ref{eq:2-channel:equation for lambda_D (normalized)}) and the condition $\tilde{\lambda}_D>0$ the dimer binding wave number is found to be
\begin{equation}
\tilde{\lambda}_D=\frac{\tilde{\mu}}{2}\left[\sqrt{\tilde{\Delta}^{2}-4\left(B-B_{\text{res}}\right)/\tilde{\mu}}-\tilde{\Delta}\right].
\end{equation}
Using the solution for $\tilde{a}$ and $\tilde{R}^\star=-\tilde{r}_e/2$ the well-known narrow resonance dimer formula
\begin{equation}
\lambda_D=\frac{\sqrt{1+4\frac{R^{\star}}{a}}-1}{2R^{\star}}
\end{equation}
is obtained.
Also this equation is far more general than the simple two-channel model.
Finally we note the connection
\begin{equation}
\Delta=\frac{\hbar^2}{m\mu R^\star}
\end{equation}
between the resonance width $\Delta=\tilde{\Delta}/k_c$ and $R^\star$.
This equation illustrates that a narrow resonance is related to a large $R^\star$.

We have shown that the bare parameters $\Lambda$ and $B_0$ are directly connected to two-body observables such as the scattering length, effective range and dimer binding energy.
For a given atomic species and scattering channel one must fix the bare parameters such that the observables are reproduced as well as possible.
Incidentally, the set $\left(\Lambda,B_0\right)$ is fully determined by $\left(R^\star,B_{\text{res}}\right)$ or $\left(\Delta,B_{\text{res}}\right)$.

\subsection{Three-Body Sector: Efimov Trimers}

Having fixed the bare parameters in the two-body sector we move on to the three-body sector with no more adjustable parameters.
To find the binding energy of Efimov trimers we search for a negative energy solution $E=-\hbar^2\lambda_T^2/m$, where $\lambda_T>\max(0,\lambda_D)$, of the Schr{\"o}dinger equation $(\hat{H}-E)|\psi_{3B}\rangle$ with
\begin{multline}
|\psi_{3B}\rangle=\int\frac{d^{3}k}{\left(2\pi\right)^{3}}\beta_{\vec{k}}\hat{b}_{\vec{k}}^{\dagger}\hat{a}_{-\vec{k}}^{\dagger}|0\rangle \\
+\int\frac{d^{3}k}{\left(2\pi\right)^{3}}\int\frac{d^{3}q}{\left(2\pi\right)^{3}}\alpha_{\vec{k},\vec{q}}\hat{a}_{\vec{q}+\frac{\vec{k}}{2}}^{\dagger}\hat{a}_{-\vec{q}+\frac{\vec{k}}{2}}^{\dagger}\hat{a}_{-\vec{k}}^{\dagger}|0\rangle.
\end{multline}
Also here we work in the center-of-mass frame and have chosen $\vec{k}$ and $\vec{q}$ to be a set of Jacobi momenta.
The Schr{\"o}dinger equation leads to the coupled equations
\begin{equation}
\left(\frac{\hbar^{2}q^{2}}{m}+\frac{3}{4}\frac{\hbar^{2}k^{2}}{m}-E\right)\alpha_{\vec{k},\vec{q}}+\Lambda\beta_{\vec{k}}=0
\end{equation}
\begin{multline}
\left(\frac{3}{4}\frac{\hbar^{2}k^{2}}{m}+E_{b}-E\right)\beta_{\vec{k}} \\
+2\Lambda\int\frac{d^{3}q}{\left(2\pi\right)^{3}}\left(\alpha_{\vec{k},\vec{q}}+2\alpha_{\vec{q}-\frac{\vec{k}}{2},-\frac{\vec{q}}{2}-\frac{3\vec{k}}{4}}\right)=0
\end{multline}
One eliminates $\alpha_{\vec{k},\vec{q}}$ from the first and plugs it into the second upon which the first integral is computed by introducing a momentum cut-off $k_c$ as in the two-body sector.
After renormalizing with respect to $k_c$, the expressions for $\tilde{a}$ and $\tilde{R}^\star$ are substituted and one obtains
\begin{multline}
\left[\sqrt{\frac{3k^2}{4}+\lambda_T^2}-\frac{1}{a}+R^\star\left(\frac{3k^2}{4}+\lambda_T^2\right)\right]\psi(k) \\
-\frac{2}{\pi}\int_0^\infty dq\ln\left(\frac{k^2+kq+q^2+\lambda_T^2}{k^2-kq+q^2+\lambda_T^2}\right)\psi(q)=0,
\label{eq:2-channel:equation for lambda_T with k,q}
\end{multline}
where $\psi(k)=k\beta_k$ and all factors of $k_c$ have canceled.
We have eliminated the magnetic field dependence (via the substitution of $a$) and can directly compute the scattering length dependence of $\lambda_T$.
We stress that the parameter $R^\star=-r_e/2$ was fixed in the two-body sector [see Eq.~(\ref{eq:2-channel:effective range is proportional to 1/Lambda^2})].
To solve Eq.~(\ref{eq:2-channel:equation for lambda_T with k,q}) we first switch variables via
\begin{equation}
k=\frac{2}{\sqrt{3}}\lambda_T\sinh(\xi) ,\quad
q=\frac{2}{\sqrt{3}}\lambda_T\sinh(\xi^\prime)
\label{eq:substitution from k to xi}
\end{equation}
and rescale $\psi(k)\rightarrow\psi(k)/\cosh(\xi)$.
If we limit ourselves to odd solutions $\psi(k)$ the lower integration limit may be extended to $-\infty$ provided we divide the entire integral by $2$.
One obtains
\begin{multline}
\left[1-\frac{1}{a\lambda_T\cosh\xi}+R^{\star}\lambda_T\cosh\xi\right]\psi\left(\xi\right) \\
-\frac{4}{\sqrt{3}\pi}\int_{-\infty}^{\infty} d\xi^{\prime}\ln\left(\frac{e^{2\left(\xi-\xi^{\prime}\right)}+e^{\xi-\xi^{\prime}}+1}{e^{2\left(\xi-\xi^{\prime}\right)}-e^{\xi-\xi^{\prime}}+1}\right)\psi\left(\xi^{\prime}\right)=0.
\end{multline}
By introducing $\int d\xi^\prime \delta(\xi-\xi^\prime)$ in the first term it is included into the integral.
One finally finds
\begin{equation}
\int_{-\infty}^{\infty}d\xi^\prime \mathcal{M}_{\lambda_T}\left(\xi,\xi^\prime\right)\psi\left(\xi^\prime\right)=0
\label{eq:2-channel:int M psi = 0 equation}
\end{equation}
with $\psi\left(\xi\right)$ an odd function of $\xi$ and
\begin{multline}
\mathcal{M}_{\lambda_T}\left(\xi,\xi^\prime\right)=\\
\delta\left(\xi-\xi^\prime\right)\left[1-\frac{1}{a\lambda_T\cosh\xi^\prime}+R^{\star}\lambda_T\cosh\xi^\prime\right] \\
-\frac{4}{\sqrt{3}\pi}\ln\left(\frac{e^{2\left(\xi-\xi^{\prime}\right)}+e^{\xi-\xi^{\prime}}+1}{e^{2\left(\xi-\xi^{\prime}\right)}-e^{\xi-\xi^{\prime}}+1}\right).
\label{eq:2-channel:matrix with xi and xi'}
\end{multline}
A non-trivial solution is obtained for $\det \mathcal{M}_{\lambda_T}\left(\xi,\xi^\prime\right)=0$ which constitutes is a closed equation for $\lambda_T$ and can be solved numerically by discretizing $\xi$ and $\xi^\prime$.
One searches for the value $\lambda_T=\lambda_T^{(\text{sol})}$ for which $\det \mathcal{M}_{\lambda_T}\left(\xi,\xi^\prime\right)$ changes sign as $\lambda_T$ is varied through $\lambda_T^{(\text{sol})}$.
Note that many solutions $\lambda_T^{(\text{sol})}$ exist for a given $a$ and $R^\star$.
To verify ones solution, the obtained value $\lambda_T^{(\text{sol})}$ is plugged back into $\mathcal{M}_{\lambda_T=\lambda_T^{(\text{sol})}}\left(\xi,\xi^\prime\right)$ and its eigen values and eigen functions $\psi(\xi)$ are computed.
One of the eigen values must be equal to zero (within machine precision) and its associated eigen function must be odd (odd number of nodes).
The value of $\lambda_T^{(\text{sol})}$ for which $\psi(\xi)$ has one node is the ground state.
If $\psi(\xi)$ has three nodes it is the first excited state and so on.
This formalism was used to produce the dashed curves in Fig.~\ref{fig:three-body}.

\section{Details on the Derivation of the Three-Channel Model}
\label{app:3-channel}

\subsection{Two-Body Sector}
\label{app:3-channel:2-body}

After plugging the two-body wave function~(\ref{eq:3-channel:two-body wave function}) into the Schr{\"o}dinger equation one obtains the following system of coupled equations:
\begin{subequations}
\begin{equation}
\left(\frac{\hbar^{2}k^{2}}{m}-E\right)\alpha_{\vec{k}}+\Lambda_{1}\beta+\Lambda_{2}\gamma=0
\end{equation}
\begin{equation}
\left(E_{b,1}-E\right)\beta+2\Lambda_{1}\int\frac{d^{3}q}{\left(2\pi\right)^{3}}\alpha_{\vec{q}}+\Lambda_{12}\gamma=0
\end{equation}
\begin{equation}
\left(E_{b,2}-E\right)\gamma+2\Lambda_{2}\int\frac{d^{3}q}{\left(2\pi\right)^{3}}\alpha_{\vec{q}}+\Lambda_{12}\beta=0
\end{equation}
\label{eq:3-channel:two-body sector three coupled equations}
\end{subequations}
From the expression for $\alpha_{\vec{k}}$ as determined from the first equation with $E=\hbar^2k_0^2/m$ and the scattering problem's Green's function [Eq.~(\ref{eq:Green's function for scattering problem})] one finds the scattering amplitude to be
\begin{equation}
f_{k_0}=-\frac{m}{4\pi\hbar^2}\left(\Lambda_{1}\beta+\Lambda_{2}\gamma\right).
\end{equation}
Then, by plugging $\alpha_{\vec{k}}$ into the second and third equation one obtains Eqs.~(\ref{eq:3-channel:equations for beta and gamma (normalized)}) for $E=\hbar^2k_0^2/m>0$, and Eqs.~(\ref{eq:3-channel:equations for lambda_D (normalized)}) for $E=\hbar^2\lambda_D^2/m<0$.
When solving the integrals in the second and third equation of~(\ref{eq:3-channel:two-body sector three coupled equations}) one must introduce a high momentum cut-off $k_c$ to prevent the divergence with respect to which the various parameters, amplitudes and variables are renormalized.
Note that Eqs.~(\ref{eq:3-channel:equations for beta and gamma (normalized)}) and~(\ref{eq:3-channel:equations for lambda_D (normalized)}) are the analogue of~(\ref{eq:2-channel:equation for beta (normalized)}) and~(\ref{eq:2-channel:equation for lambda_D (normalized)}) in the the two-channel model.

\subsection{Relating the Bare Parameters to Observable Parameters}
\label{app:3-channel:2-body analytic expressions}

In terms of the bare parameters the observable parameters are given via
\begin{subequations}
\begin{equation}
\tilde{\Delta}_{1}=-\frac{L_{s}^{2}}{4\pi}+\frac{L_{d}^{2}\left(B_{1}-B_{2}\right)+L_{p}^{4}}{4\pi R_{1}}
\end{equation}
\begin{equation}
\tilde{\Delta}_{2}=-\frac{L_{s}^{2}}{4\pi}-\frac{L_{d}^{2}\left(B_{1}-B_{2}\right)+L_{p}^{4}}{4\pi R_{1}}
\end{equation}
\begin{equation}
B^{(\text{res})}_1=\frac{B_{1}+B_{2}}{2}-\frac{L_{s}^{2}}{2\pi^{2}}-\frac{R_{2}}{2\pi^{2}}
\end{equation}
\begin{equation}
B^{(\text{res})}_2=\frac{B_{1}+B_{2}}{2}-\frac{L_{s}^{2}}{2\pi^{2}}+\frac{R_{2}}{2\pi^{2}}
\end{equation}
\end{subequations}
where we have defined
\begin{subequations}
\begin{multline}
R_{1}=\bigg\{ 2\left(L_{1}L_{2}-2\pi^{2}L_{12}\right)^{2}+2L_{1}^{2}L_{2}^{2}-4\pi^{4}L_{12}^{2} \\
+\left[\pi^{2}(B_{1}-B_{2})-\left(L_{1}^{2}-L_{2}^{2}\right)\right]^{2} \bigg\}^{1/2}
\end{multline}
\begin{multline}
R_{2}=\bigg\{ 4\pi^{4}L_{12}^{2}-8\pi^{2}L_{1}L_{2}L_{12}+L_{s}^{4} \\
-2L_{d}^{2}\left(B_{1}-B_{2}\right)+\pi^{4}\left(B_{1}-B_{2}\right)^{2} \bigg\}^{1/2}
\end{multline}
\end{subequations}
as well as $L_{s}^{2}=L_{1}^{2}+L_{2}^{2}$, $L_{d}^{2}=\pi^{2}\left(L_{1}^{2}-L_{2}^{2}\right)$ and 
$L_{p}^{4}=4\pi^{2}L_{1}L_{2}L_{12}-2L_{1}^{2}L_{2}^{2}-L_{1}^{4}-L_{2}^{4}$.
The $L$-parameters correspond to the $\Lambda$-parameters scaled by the magnetic moment, so: $L_{1,2}=\tilde{\Lambda}_{1,2}/\sqrt{\tilde{\mu}_{1,2}}$ and $L_{12}=\tilde{\Lambda}_{12}/\sqrt{\tilde{\mu}_{1}\tilde{\mu}_{2}}$.

Note that the functions produce $\tilde{\Delta}_{1,2}$ which are related to $\Delta_{1,2}$ in Table~\ref{tb:model systems parameters} via $\tilde{\Delta}_{1,2}/k_c=\Delta_{1,2}$.
Hence, $k_c$ must be fixed before solving the equations for the bare parameters.

{\yy
\subsection{Relation of Eq.~(\ref{eq:3-channel:Feshbach resonance formula}) to Eq. (26) in Ref.~\cite{Jachymski13}}
\label{app:3-channel:compare to Jachymski}

For the case of two resonances and vanishing background scattering, Eq. (26) of Ref.~\cite{Jachymski13} can be written as
\begin{widetext}
\begin{equation}
a(B)= 
\frac{\Delta_{1}\left(B_{2}-B\right)+\Delta_{2}\left(B_{1}+B\right)}{\left(B_{1}-B\right)\left(B_{2}-B\right)+\left(B_{1}-B\right)\delta B_{2}+\left(B_{2}-B\right)\delta B_{1}}.
\end{equation}
\end{widetext}
To obtain this form one must take $a_{bg}\rightarrow0$ and $\Delta_{1,2}\rightarrow\infty$ while keeping their product (which we call $\Delta_{1,2}$) finite.
Furthermore, our expression for the scattering length can be written in the form
\begin{widetext}
\begin{equation}
a(B)= 
\frac{-\frac{L_{1}^{2}}{2\pi}\left(B_{2}-B\right)-\frac{L_{2}^{2}}{2\pi}\left(B_{1}-B\right)+\frac{L_1L_2L_{12}^4}{\pi}}{\left(B_{1}-B\right)\left(B_{2}-B\right)-\frac{L_{2}^{2}}{\pi^{2}}\left(B_{1}-B\right)-\frac{L_{1}^{2}}{\pi^{2}}\left(B_{2}-B\right)+2\frac{L_1L_2L_{12}}{\pi^{2}}-L_{12}^{2}}.
\end{equation}
\end{widetext}
For $L_{12}=0$ one recognizes the same combinations of $\left(B_{i}-B\right)$ in the two expressions.
Our model is thus consistent with the previously used framework of multi-channel quantum defect theory.
Note that the apparent connection between $\Delta_i$ and $\delta B_i$ suggested by comparison of the two equaitons does not exist in Ref.~\cite{Jachymski13}.
This is due to the presence of the van der Waals potential in Ref.~\cite{Jachymski13}; see Eq.~(22) there.
}

\subsection{Three-Body Sector}
\label{app:3-channel:3-body}

From the Schr{\"o}dinger equation $(\hat{H}-E_T)|\psi_{3B}\rangle=0$, where $|\psi_{3B}\rangle$ is given in Eq.~(\ref{eq:3-channel:three-body wave function}), one obtains three coupled integral equations:
\begin{widetext}
\begin{subequations}
\begin{equation}
\left(\frac{\hbar^{2}q^{2}}{m}+\frac{3}{4}\frac{\hbar^{2}k^{2}}{m}-E_T\right)\alpha_{\vec{k},\vec{q}}+\Lambda_{1}\beta_{\vec{k}}+\Lambda_{2}\gamma_{\vec{k}}=0
\end{equation}
\begin{equation}
\left(\frac{3}{4}\frac{\hbar^{2}k^{2}}{m}+E_{b,1}-E_T\right)\beta_{\vec{k}}+\Lambda_{12}\gamma_{\vec{k}}+2\Lambda_{1}\int\frac{d^{3}q}{\left(2\pi\right)^{3}}\left[\alpha_{\vec{k},\vec{q}}+2\alpha_{\vec{q}-\frac{\vec{k}}{2},-\frac{\vec{q}}{2}-\frac{3\vec{k}}{4}}\right]=0
\end{equation}
\begin{equation}
\left(\frac{3}{4}\frac{\hbar^{2}k^{2}}{m}+E_{b,2}-E_T\right)\gamma_{\vec{k}}+\Lambda_{12}\beta_{\vec{k}}+2\Lambda_{2}\int\frac{d^{3}q}{\left(2\pi\right)^{3}}\left[\alpha_{\vec{k},\vec{q}}+2\alpha_{\vec{q}-\frac{\vec{k}}{2},-\frac{\vec{q}}{2}-\frac{3\vec{k}}{4}}\right]=0
\end{equation}
\end{subequations}
\end{widetext}
The free particle amplitude $\alpha_{\vec{k},\vec{q}}$ is eliminated from the first equation and plugged into the second and third.
The first of the two integrals can be solved as in the two-body sector by introducing a high momentum cut-off $k_c$ with which the coupling constants are renormalized according to $\tilde{\Lambda}_1=\Lambda_1 k_{c}^{3/2}/E_{c}$, $\tilde{\Lambda}_2=\Lambda_2 k_{c}^{3/2}/E_{c}$ and $\tilde{\Lambda}_{12}=\Lambda_{12},E_{c}$, and the amplitudes according to $\tilde{\beta}=\beta k_{c}^{3/2}$ and $\tilde{\gamma}=\gamma k_{c}^{3/2}$.
The renormalized magnetic moment is $\tilde{\mu}_i=\mu_i/E_{c}$ and all momenta are $\tilde{k}=k/k_c$.
In addition one uses the $s$-wave property that $\beta_{\vec{k}}=\beta_{k}$ and $\gamma_{\vec{k}}=\gamma_{k} $ are spherically symmetric.
The equations are then
\begin{widetext}
\begin{subequations}
\begin{multline}
\left(\frac{3}{4}\tilde{k}^{2}+\tilde{\lambda}_T^{2}+\tilde{\mu}_{1}\left(B_{1}-B\right)\right)\tilde{\beta}_{k}+\tilde{\Lambda}_{12}\tilde{\gamma}_{k}
-\frac{\tilde{\Lambda}_{1}}{\pi^{2}}\left(1-\frac{\pi}{2}\sqrt{\frac{3}{4}\tilde{k}^{2}+\tilde{\lambda}_T^{2}}\right)\left(\tilde{\Lambda}_{1}\tilde{\beta}_{k}+\tilde{\Lambda}_{2}\tilde{\gamma}_{k}\right) \\
-\frac{\tilde{\Lambda}_{1}}{\pi^{2}}\int_0^\infty d\tilde{q}\frac{\tilde{q}}{\tilde{k}}\ln\left(\frac{\tilde{q}^{2}+\tilde{q}\tilde{k}+\tilde{k}^{2}+\tilde{\lambda}_T^{2}}{\tilde{q}^{2}-\tilde{q}\tilde{k}+\tilde{k}^{2}+\tilde{\lambda}_T^{2}}\right)\left(\tilde{\Lambda}_{1}\tilde{\beta}_{q}+\tilde{\Lambda}_{2}\tilde{\gamma}_{q}\right)=0
\end{multline}
\begin{multline}
\left(\frac{3}{4}\tilde{k}^{2}+\tilde{\lambda}_T^{2}+\tilde{\mu}_{2}\left(B_{2}-B\right)\right)\tilde{\gamma}_{k}+\tilde{\Lambda}_{12}\tilde{\beta}_{k}
-\frac{\tilde{\Lambda}_{2}}{\pi^{2}}\left(1-\frac{\pi}{2}\sqrt{\frac{3}{4}\tilde{k}^{2}+\tilde{\lambda}_T^{2}}\right)\left(\tilde{\Lambda}_{1}\tilde{\beta}_{k}+\tilde{\Lambda}_{2}\tilde{\gamma}_{k}\right) \\
-\frac{\tilde{\Lambda}_{2}}{\pi^{2}}\int_0^\infty d\tilde{q}\frac{\tilde{q}}{\tilde{k}}\ln\left(\frac{\tilde{q}^{2}+\tilde{q}\tilde{k}+\tilde{k}^{2}+\tilde{\lambda}_T^{2}}{\tilde{q}^{2}-\tilde{q}\tilde{k}+\tilde{k}^{2}+\tilde{\lambda}_T^{2}}\right)\left(\tilde{\Lambda}_{1}\tilde{\beta}_{q}+\tilde{\Lambda}_{2}\tilde{\gamma}_{q}\right)=0
\end{multline}
\label{eq:3-channel:two coupled equations in terms of k,q}
\end{subequations}
\end{widetext}
From here one proceeds equivalent to the two-channel model.
The amplitudes are rescaled according to $\tilde{\beta}_{k}\rightarrow\tilde{\beta}_{k}/\tilde{k}$ and $\tilde{\gamma}_{k}\rightarrow\tilde{\gamma}_{k}/\tilde{k}$.
We then switch from $(\tilde{k},\tilde{q})$ to $(\xi,\xi^\prime)$ using Eq.~(\ref{eq:substitution from k to xi}) and rescale once more according to $\tilde{\beta}_{\xi}\rightarrow\tilde{\beta}_{\xi}/\cosh\xi$ and $\tilde{\gamma}_{\xi}\rightarrow\tilde{\gamma}_{\xi}/\cosh\xi$.
Finally, the lower integration limit is extended to $-\infty$ and the integral divided by $2$.
The amplitudes must thus be odd functions of $\xi$.
One obtains
\begin{widetext}
\begin{subequations}
\begin{equation}
\int_{-\infty}^{\infty}d\xi^{\prime}\bigg[\left\{ F_{1}\left(\xi^{\prime}\right)\delta\left(\xi-\xi^{\prime}\right)-\tilde{\Lambda}_{1}^{2}L\left(\xi,\xi^{\prime}\right)\right\} \tilde{\beta}_{\xi^{\prime}} \\
+\left\{ H\left(\xi^\prime\right)\delta\left(\xi-\xi^{\prime}\right)-\tilde{\Lambda}_{1}\tilde{\Lambda}_{2}L\left(\xi,\xi^{\prime}\right)\right\} \tilde{\gamma}_{\xi^{\prime}}\bigg]=0
\end{equation}
\begin{equation}
\int_{-\infty}^{\infty}d\xi^{\prime}\bigg[\left\{ H\left(\xi^{\prime}\right)\delta\left(\xi-\xi^{\prime}\right)-\tilde{\Lambda}_{1}\tilde{\Lambda}_{2}L\left(\xi,\xi^{\prime}\right)\right\} \tilde{\beta}_{\xi^{\prime}}
+\left\{ F_{2}\left(\xi^{\prime}\right)\delta\left(\xi-\xi^{\prime}\right)-\tilde{\Lambda}_{2}^{2}L\left(\xi,\xi^{\prime}\right)\right\} \tilde{\gamma}_{\xi^{\prime}}\bigg]=0
\end{equation}
\end{subequations}
\end{widetext}
where
\begin{subequations}
\begin{equation}
F_i(\xi)=f_i(\xi)-\tilde{\Lambda}_i^2g(\xi)
\end{equation}
\begin{equation}
H(\xi)=h(\xi)-\tilde{\Lambda}_1\tilde{\Lambda}_2g(\xi)
\end{equation}
\end{subequations}
As discussed in the main text, these two coupled integral equations can be written in matrix form and the condition of vanishing determinant gives a closed equation for $\lambda_T$.
In order to solve Eq.~(\ref{eq:3-channel:equantion for lambda_T, determinant = 0}), each block $\mathcal{M}_{ij}$ is written as a $n\times n$ matrix by discretizing $\xi$ and $\xi^\prime$ in the interval $\left[-\xi_m,\xi_m\right]$ and step size $d\xi=2\xi_m/(n-1)$.
The total matrix thus has dimensions $2n\times2n$ and its determinant is found numerically.
In Fig.~\ref{fig:three-body}, we used $\xi_m=10.02$ (to avoid {\yy the singularity at} $\xi=0$) and $n=50$.

\subsection{Equation for Eigenfunction Ratio}
\label{app:3-channel:3-body eigen function ratio}

The first and second equations of~(\ref{eq:3-channel:two coupled equations in terms of k,q}) are multiplied by $\tilde{\Lambda}_2$ and $\tilde{\Lambda}_1$ respectively.
Subtracting the second from the first then gives
\begin{multline}
\left[\tilde{\Lambda}_{2}\left(\frac{3}{4}\tilde{k}^{2}+\tilde{\lambda}_{T}^{2}+\tilde{\mu}_{1}\left(B_{1}-B\right)\right)-\tilde{\Lambda}_{1}\tilde{\Lambda}_{12}\right]\tilde{\beta}_{k} \\
=\left[\tilde{\Lambda}_{1}\left(\frac{3}{4}\tilde{k}^{2}+\tilde{\lambda}_{T}^{2}+\tilde{\mu}_{2}\left(B_{2}-B\right)\right)-\tilde{\Lambda}_{2}\tilde{\Lambda}_{12}\right]\tilde{\gamma}_{k}
\end{multline}
which leads directly to expression~(\ref{eq:3-channel:ratio chi_k in terms of bare parameters}) for the ratio $\chi_k=\tilde{\beta}_k/\tilde{\gamma}_k$.
Looking for zero-crossings at positive $k$, so $\chi_{k}=0$ for $k\geq0$, leads to condition~(\ref{eq:3-channel:inequality for critical field (1+1) to (3+1) nodes}).

\section{Two-Body Sector of the $bb$-channel of $^7$Li}
\label{app:Li-7}

\begin{figure}
\centering
\includegraphics[width=0.9\linewidth]{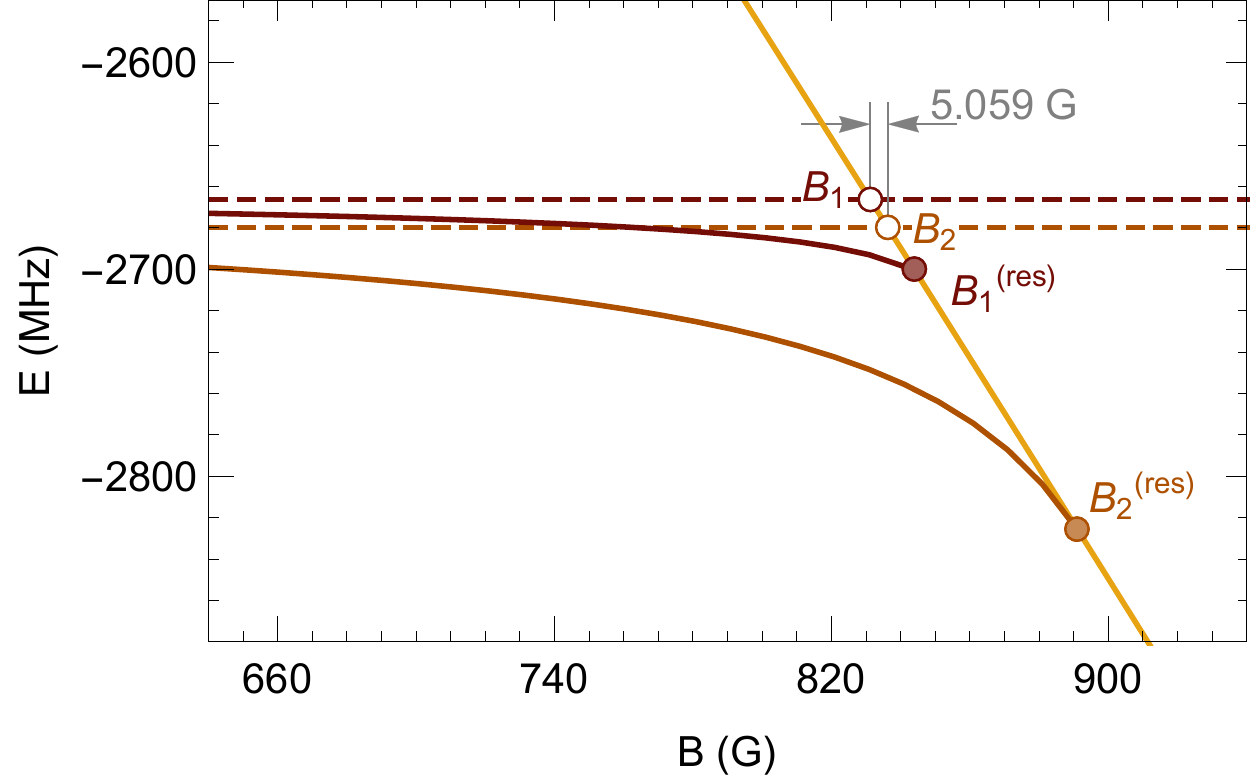}
\caption{\label{fig:Li-7 coupled channels}
Magnetic field dependence of the three relevant channels as obtained from coupled-channels calculations.
The free atoms (yellow) experience a linear Zeeman shift.
The bare molecules (dashed light and dark brown) are spin singlets and hence magnetic field independent.
The interaction terms of the three-channel model give rise to the physical dimers (solid light and dark brown).
}
\end{figure}

\begin{figure}
\centering
\includegraphics[width=1.\linewidth]{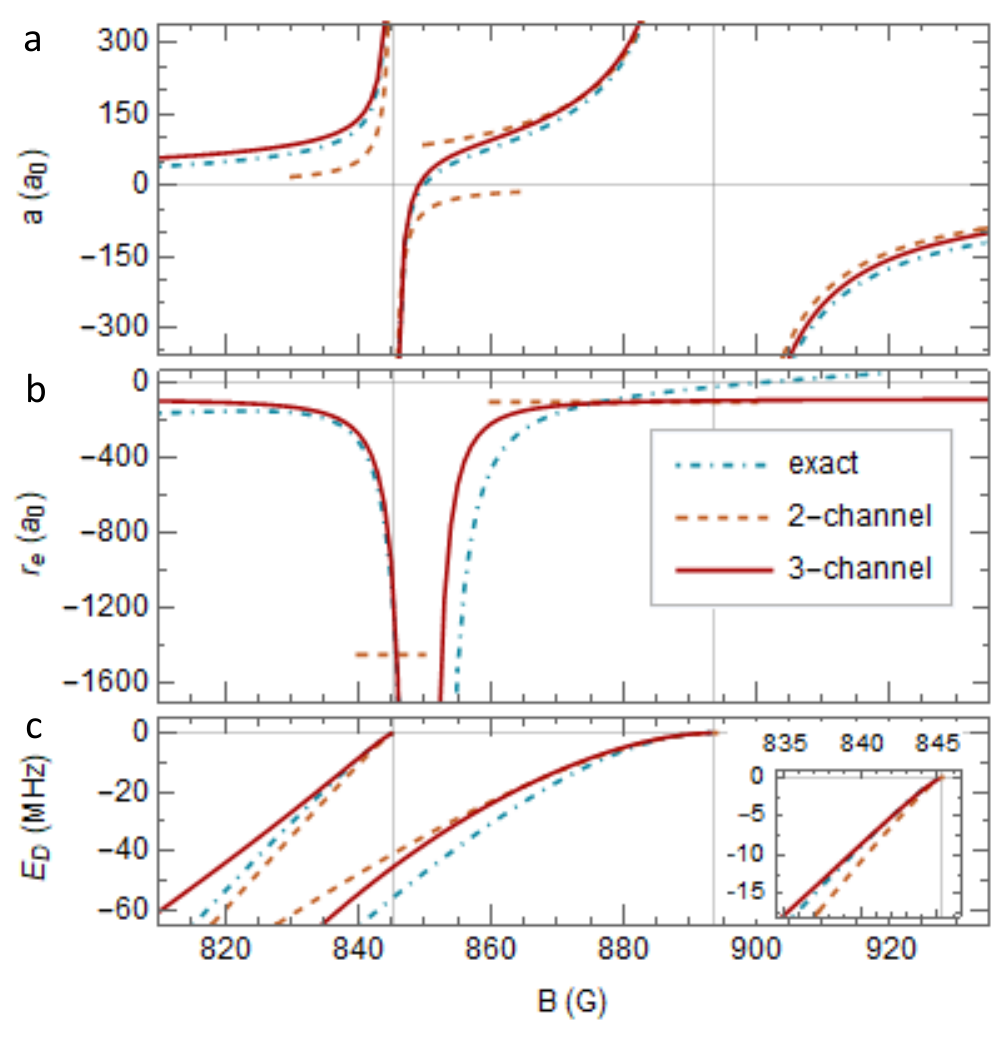}
\caption{\label{fig:Li-7 two-body sector}
The (a) scattering length (b) effective range and (c) dimer binding energy of the $bb$-channel in $^7$Li as obtained from the three-channel model are compared to the exact coupled-channels result and the individual two-channel model treatments.
}
\end{figure}

Here the two-body sector equations of the three-channel model are applied to the $bb$-channel of $^7$Li.
Fig.~\ref{fig:Li-7 coupled channels} shows coupled-channels calculations of the relevant energy levels (channels).
Both bare molecular states are spin singlets and therefore have no magnetic moment, i.e. they are magnetic field independent.
Since the two resonances occur at high magnetic field the Zeeman shift is linear in the region of interest.
The two free atoms are essentially a spin triplet and have a combined magnetic moment of $\mu=-2.66$~MHz/G, which is close to $-2\mu_B$ of a full triplet, where $\mu_B=1.4$~MHz/G is the Bohr magneton.

We note that the coupled-channels data shown in Fig.~\ref{fig:Li-7 coupled channels} illustrates the ingredients and phenomenology of the three-channel model.
The two closed channels of $\hat{H}_0$ (straight dashed lines) intersect the open channel (straight yellow line) with relative slope $\mu=\mu_1=\mu_2$ at the bare resonance positions $B_i$.
The energy differences between the two dashed and the yellow solid line are the bare binding energies $E_{b,i}$ and are negative for $B<B_i$.
The atom-molecule couplings $\hat{H}_{i}$ and the inter-molecular coupling $\hat{H}_{12}$ shift the actual resonance position to lower energies and higher magnetic fields and alter the dimer binding energy as illustrated.

We go through the following three steps to fix the bare parameters.
(1) Fitting Eq.~(\ref{eq:3-channel:Feshbach resonance formula}) to coupled-channels data of the scattering length gives $\left(B_1^{(\text{res})},B_2^{(\text{res})}\right)=\left(845.3,893.7\right)$ G, which are in perfect agreement with the coupled-channels resonance positions, and $\left(\Delta_1,\Delta_2\right)/a_0=\left(342.37,3996.93\right)$ G.
(2) To approximately satisfy $k_c\sim1/r_{vdW}$ we choose $k_c=1/(60a_0)$.
(3) Using $B_2-B_1=5.029$ G (Fig.~\ref{fig:Li-7 coupled channels}) as the additional constraint (as suggested in Sec.~\ref{sec:Lambda_bb}) and the analytic expressions from Appendix~\ref{app:3-channel:2-body analytic expressions} we find $\left(\tilde{\Lambda}_1,\tilde{\Lambda}_2,\tilde{\Lambda}_{12}\right)=\left(0.561,2.849,0.243\right)$ and $\left(B_1,B_2\right)=\left(844,849.1\right)$ G.

The computed scattering length, effective range and dimer binding energy are shown in Fig.~\ref{fig:Li-7 two-body sector}.
The coupled-channels scattering length is consistently below the three-channel result.
This is to be expected since the coupled-channels calculation includes background scattering in the open channel and the shift agrees with its magnitude $a_{bg}=-18.42a_0$.
The two-channel model captures the scattering length only in a small window around the resonance positions which is by-far narrower than the resonance widths $\Delta B$.
Moreover, since both resonances are treated individually, it is not capable of reproducing the zero-crossing.
The fact that the three-channel model {\it does} capture the zero-crossing can be appreciated by looking at the effective range.
In Fig.~\ref{fig:Li-7 two-body sector}(b) the effective range is seen to diverge at the scattering length zero-crossing for both the coupled-channels and three-channel calculation.
According to the two-channel model however, the effective range is magnetic field independent and agrees with the three-channel model only at the resonance positions.
The difference of the coupled-channels effective range at the higher resonance is due to van der Waals physics~\cite{Gao11,Werner12} not taken into account by either of the models.
Finally, Fig.~\ref{fig:Li-7 two-body sector}(c) and its Inset show that also the dimer binding energy is captured better by an overlapping resonance theory then by an individual treatment especially in the shallow binding regime.
At larger binding energy the inter-molecular coupling manifests itself as level repulsion.
However, as discussed in Sec.~\ref{sec:Lambda_bb}, the level repulsion is not related to the numerical value assigned to $\tilde{\Lambda}_{12}$ -- which here was constrainded by $B_2-B_1=5.029$ G -- but to the intrinsic cross-talk of the two dimers.


\begin{thebibliography}{10}
\bibitem{broadVSnarrow}
The meaning of ``broad'' and ``narrow'' in the classification of Feshbach
  resonances refers to the value of $s_{res}$. It is not equivalent to the
  parameter $\Delta B$ in $a=a_{bg}\left[1-\Delta B/\left(B-B_0\right)\right]$
  which is usually termed ``resonance width'' and signifies the distance (in
  units of magnetic field) between the resonance position $B_0$ and the
  zero-crossing. The $s_{res}$ parameter is more general; in addition to
  $\Delta B$ it also depends on $a_{bg}$ whose value determines the position of
  the zero crossing. In fact, up to numerical coefficients on the order of
  unity and factors of $\hbar$ and mass, $s_{res}$ is proportional to
  $a_{bg}r_{vdW}\mu\Delta B$~\cite{Chin10}.
\bibitem{DeltaVSDeltaB}
Since our model does not include background scattering (so $a_{bg}=0$) the
  formula $a=a_{bg}\left[1-\Delta B/\left(B-B_0\right)\right]$, or its
  extension to multiple resonances~\cite{Jachymski13}, is not applicable. The
  correct single-resonance formula is $a=a_{bg}\Delta B/\left(B-B_0\right)$,
  where $a_{bg}\rightarrow0$ and $\Delta B\rightarrow\infty$ while their
  product $\Delta=a_{bg}\Delta B$ remains constant. Thus, the width parameter
  $\Delta$ has dimensions of magnetic field times length and the renormalized
  $\tilde{\Delta}=k_c\Delta$ is in units of magnetic field.
\bibitem{dimerAtomScatteringResonances}
The trimers that are associated with the resonance at $B_1^{(\text{res})}$ and
  embedded in the dimer-atom continuum of the dimer associated with the
  resonance at $B_2^{(\text{res})}$ are manifested as scattering resonances
  between a $B_2^{(\text{res})}$-dimer and a free atom. It should therefore be
  possible to find them in a scheme similar to the computation of the
  three-body recombination rate at $E>0$ in which $a_-^{(n)}$ shows up as a
  resonance~\cite{Jona-Lasinio08}.
\end{thebibliography}
\bibliographystyle{unsrt}

\end{document}